\newcommand{\Msun}{\mbox{$M_{\odot}$}}
\newcommand{\msun}{\mbox{$M_{\odot}$}}

\newcommand{\Rgal}{\mbox{$R_{\rm Gal}$}}
\newcommand{\Vgal}{\mbox{$V_{\rm Gal}$}}

\newcommand{\Mi}{\mbox{$M_{\rm i}$}}
\newcommand{\Sn}{\mbox{$S_N$}}
\newcommand{\MH}{\mbox{$[M/H]$}}
\newcommand{\FeH}{\mbox{$[M/H]$}}
\newcommand{\Feh}{\mbox{$[M/H]$}}
\newcommand{\feh}{\mbox{$[M/H]$}}
\newcommand{\Rgcst}{\mbox{$R_{\rm GC/stars}$}}

\newcommand{\kms}{\mbox{km.s$^{-1}$}}

\documentclass{aa}  

\usepackage{graphicx, bm, amssymb, xcolor}
\usepackage{verbatim}
\usepackage[breaklinks,colorlinks,citecolor=blue]{hyperref}
\usepackage[all]{hypcap}

\usepackage{aas_macros}
\usepackage{epsfig}
\usepackage{txfonts}
\usepackage{natbib}
\bibpunct{(}{)}{;}{a}{}{,}   
\begin{document}

   \title{{\bf The difference in metallicity distribution functions of halo stars and 
globular clusters as a function of galaxy type
}}

   \subtitle{A tracer of globular cluster formation and evolution}
   
\author{H.~J.~G.~L.~M.~Lamers \inst{1}
        \and J.~M.~D.~Kruijssen\inst{2}
	\and N.~Bastian\inst{3}
        \and M.~Rejkuba\inst{4,5}
        \and M.~Hilker\inst{4,5}     
        \and M.~Kissler-Patig\inst{6}    
}

   \offprints{H.~J.~G.~L.~M. Lamers}

   \institute{Astronomical Institute Anton Pannekoek, University of Amsterdam,  Science Park 904, NL-1098XH, Amsterdam, The Netherlands\\
              \email{h.j.g.l.m.lamers@uu.nl}
             \and              
              Astronomisches Rechen-Institut, Zentrum f\"{u}r Astronomie der Universit\"{a}t Heidelberg, Monchhofstra\ss e 12-14, D-69120 Heidelberg, Germany\\
              \email{kruijssen@uni-heidelberg.de}
            \and 
              Astrophysics Research Institute, Liverpool John Moores University,
              Egerton Wharf, Birkenhead, CH41 1LD, UK.\\
	      \email{n.j.bastian@ljmu.ac.uk}
	     \and
             ESO, Karl-Schwarzschild-Strasse
              2, D-85748 Garching, Germany\\
              \email{[mrejkuba,mhilker]@eso.org}
             \and
              Excellence Cluster Universe, Boltzmannstr. 2, D-85748, Garching, Germany
             \and
             Gemini Observatory, 670 N. A'ohoku Place, Hilo, Hawaii 96720, USA,
              \email{mkissler@gemini.edu}
             }

   \date{Received 28 April 2017; Accepted 2 June 2017}
\titlerunning{Metallicity distributions of stars and clusters}


  \abstract
  { 
  Observations of globular clusters (GCs) and field stars in the halos of the giant elliptical galaxy Cen~A and the spiral galaxy M31 show a large range of cluster-to-star number ratios (or `specific frequencies'). The cluster-to-star ratio decreases with increasing metallicity by over a factor of 100--1000, at all galactocentric radii and with a slope that does not seem to depend on radius. In dwarf galaxies, the GCs are also more metal-poor than the field stars on average. These observations indicate a strong dependence of either the cluster formation efficiency and/or the cluster destruction rate on metallicity and environment.
  }
  {
  We aim to explain the observed trends by considering the various effects that influence the cluster-to-star ratio as a function of metallicity, environment and cosmological history.
  }
{ 
  We discuss the following effects that may influence the observed cluster-to-star ratio:
    (a) the formation efficiency of GCs; 
    (b) the destruction of embedded GCs by gas expulsion;
    (c) the maximum masses of GCs;
    (d) the destruction of GCs by tidal stripping, dynamical friction, and tidal shocks as a function of environment;
    (e) the hierarchical assembly of GC systems during galaxy formation and the dependence on metallicity.
}
{
    We show that both the cluster formation efficiency and the maximum cluster mass increase with metallicity, so they cannot explain the observed trend. Destruction of GCs by tidal stripping and dynamical friction destroy clusters mostly within the inner few kpc, whereas the cluster-to-star ratio trend is observed over a much larger range of galactocentric radii. We show that cluster destruction by tidal shocks from giant molecular clouds in the high-density formation environments of GCs becomes increasingly efficient towards high galaxy masses and, hence, towards high metallicities. The predicted cluster-to-star ratio decreases by a factor 100-1000 towards high metallicities and should only weakly depend on galactocentric radius due to orbital mixing during hierarchical galaxy merging, consistent with the observations.
 }
{ 
The observed, strong dependence of the cluster-to-star ratio on metallicity and the independence of its slope on galactocentric radius can be explained by cluster destruction and hierarchical galaxy growth. During galaxy assembly, GC metallicities remain a good tracer of the host galaxy masses in which the GCs formed and experienced most of their destruction. As a result, we find that the metallicity-dependence of the cluster-to-star ratio does not reflect a GC formation efficiency, but a survival fraction.
}
   \keywords{globular clusters: general;
             galaxies: star clusters: general;
             galaxies: abundances;
             galaxies: stellar content;
             galaxies: halos;
             galaxies: star formation
             }
   \maketitle
%

\section{Introduction} \label{sec:intro}

 Studies of the halo of the Milky Way and other massive galaxies show that the metallicity distribution functions (MDFs) of halo stars is very different from that of globular clusters in the same galaxy. Observations indicate that the 
cluster-to-star ratio is a strong function of metallicity 
in spiral (e.g.~\citealt{carollo10, gratton12, durrell01, durrell04, chapman06,kalirai06}) and elliptical galaxies \citep[e.g.][]{harrisharris02, harris03, beasley08}. Similar differences  have also been observed for dwarf galaxies (e.g.~\citealt{beasley08, mackey04, sharina10}) \footnote{The cluster-to-star ratio is the mass of stars in clusters compared to the mass of field stars in the same area of a galaxy.}.
 The most extreme example was found in the Fornax dSph where the stars have a broad metallicity distribution with $-2.5 < \Feh < -0.2$ with a peak at $\Feh =-1.0$ while the globular clusters are all metal-poor with $-2.6 < \Feh < -1.4$ \citep{larsen12b}. So the cluster-to-star ratio is very low at high metallicity and very high at low metallicity.

These trends are not expected if star formation and cluster formation trace each other as in nearby star-forming regions where a certain small fraction of the stars \citep[typically 5--10\%,][]{goddard10,kruijssen12d,adamo15b} are in bound clusters. If ancient clusters and stars would also trace each other, one would expect that the metallicity distributions of the old globular clusters and old stellar populations match each other (assuming no metal contamination within clusters), which is obviously not the case. One interpretation of these results is that star and cluster formation were different in the early universe, when metal-poor GCs were formed, with separate events forming the bulk of the cluster and stellar populations.  Alternatively, it is possible that evolutionary effects are responsible for the observed differences, with preferential cluster disruption or hierarchical galaxy merging influencing the ratio of clusters to field stars as a function of metallicity or location.
 
There has been much work done in the past few years attempting to include globular clusters in galaxy formation simulations in an attempt to place globular clusters in a more cosmological context and obtain an understanding of their MDFs.  One commonly adopted method is through `particle tagging', where certain particles in the simulations are treated as globular clusters.  Such simulations have provided insight into various globular cluster population properties \citep[e.g.][]{moore06,moran14}, although they generally make the assumption that field stars and globular clusters will share the same MDFs, whereas observations show variations over $\sim3$ orders of magnitude \citep[e.g.][]{kruijssen14c}.  Such assumptions lead to unphysical solutions as to when globular clusters formed (i.e.~$z>10$ for metal poor clusters, see \citealt{brodie06}, which is inconsistent with their ages, see e.g.~\citealt{forbes10}), because they can only produce sufficiently extended globular cluster galactocentric radial distributions at extremely high redshift.  Additionally, such simulations, often done in post-processing, do not follow the tidal history of the globular clusters meaning that the cluster dissolution cannot be taken into account in the necessary detail. \citet{tonini13} got around this problem by matching the cluster-to-star ratio (or specific frequency) of each galaxy (in post-processing) to observations of present-day galaxies, with the aim of reproducing the MDFs of globular clusters. However, the origin of the adopted specific frequency cannot be tested in this way. When aiming to understand the origin of the different MDFs of globular clusters and field stars, which is equivalent to studying the cluster-to-star ratio as a function of metallicity, one must instead employ fully self-consistent models for the formation and evolution of globular cluster in galaxy formation simulations, taking into account the impact of the environment on the cluster formation efficiency and cluster mass loss.

In this paper, we aim to formulate the problem of the different MDFs of globular clusters and field stars and identify the most promising responsible physical mechanisms, particularly with an eye on the upcoming generation of self-consistent models of globular cluster formation during galaxy formation (e.g.~Pfeffer et al.~in prep., Kruijssen et al.~in prep.).
We do this by describing the observational constraints and studying the different effects that influence both the formation and the disruption of star clusters, and may have played a role in producing different MDFs of globular clusters and halo stars within galaxies.
\begin{enumerate}
\item The first is that the fraction of stars formed in bound clusters ($\Gamma$) may depend on metallicity and environment. 
\item The second, intimately related to the first, is that some internal or external process may preferentially destroy young clusters in a metallicity dependent way.
\item The third is that the maximum cluster mass may depend on metallicity such that massive globular clusters are less likely to form in metal-rich galaxies.
\item The fourth is environmentally dependent cluster disruption, where clusters may be destroyed more rapidly in strong and/or time-varying tidal fields, and most strongly towards high metallicities (e.g., near the centres of galaxies or in gas-rich environments). 
\item The fifth is dynamical friction, which may force massive clusters to spiral-in to the centres of galaxies.
\item The sixth is any combination of the first five, considered in the framework of hierarchical galaxy formation. 
\end{enumerate}

The goal of the present work is to investigate why the cluster and star populations have different MDFs within galaxies. 
The paper is organized as follows.
 In \S~\ref{sec:obs} we describe the relevant observations of the distributions (spatial and metallicity) of old stars and globular clusters in halos of spiral, elliptical and dwarf galaxies.  
In \S~\ref{sec:obssum} we briefly summarize these observations and mention the unexplained problems.
In \S~\ref{sec:form} we discuss the formation of clusters of different metallicities in different environments.
In \S~\ref{sec:destruc} the different mechanisms that may destroy the clusters are discussed.
In \S~\ref{sec:cosmology} the capture of clusters from stripped dwarf galaxies are studied.
The conclusions and discussion are presented in \S\ref{sec:conclusions}.

\section{Observed metallicity distribution functions (MDF) of clusters and stars} \label{sec:obs}

The study of the MDF of stars in individual 
galaxy halos is possible through resolved stellar photometry, but is 
limited to galaxies in the Local Group and nearby groups, up to about 10~Mpc. 
We will use the detailed properties of stars and GCs in Cen A, M31 and the Fornax dSphs as representative examples for ellipticals, spirals and dwarf galaxies.

\subsection{Large spiral and elliptical galaxies} \label{sec:obs:spirals}

\subsubsection{The MDF of halo stars} \label{sec:2.1.1}

The stars in the halos of large spiral and elliptical galaxies show similarities in their metallicity distributions:
a high metallicity component with a peak at $[M/H] \simeq -0.5~ {\rm to}~ -1.5$ and a lower metallicity component or a tail in more massive elliptical galaxies, with a broader peak at $[M/H] \simeq  -1.5~ {\rm to}~-2.0$, e.g.:  
{\bf MW} : \citep{ryan91,an12};
{\bf M31}: \citep{durrell01,durrell04,chapman06,kalirai06, gilbert14, ibata14, gregersen15}; 
{\bf M81}: \citep{durrell10};
{\bf NGC~891}: \citep{mouhcine07,rejkuba09}; 
{\bf NGC~2403}: \citep{barker12}; 
{\bf Sombrero}: \citep{mould10}; 
 other nearby spirals \citep[e.g.][]{mouhcine05,mouhcine06}; 
{\bf NGC~5128}: \citep{harris99, harris00, harrisharris02,rejkuba05, crnojevic13, rejkuba14};
{\bf NGC~3377}: \citep{harris07b}
{\bf NGC~3379}: \citep{harris07a,leejang16}
{\bf NGC~3115}: \citep{peacock15}.

The metal-rich component is concentrated more towards the inner regions
and the metal-poor component becomes significant only in the outer regions. 
The metallicity of both the metal-rich and metal-poor 
components increase with increasing galaxy luminosity \citep{mouhcine06} and the stellar halo metallicity scales with  
stellar halo mass \citep{harmsen17} with a slope that is broadly similar to the stellar mass versus stellar metallicity relation for local galaxies \citep{gallazzi05, kirby13}.  \citet{forbes97} and \citet{mouhcine06} suggested that the metal-poor 
component is due to the tidal disruption of dwarf-like objects whereas the metal-rich 
population is related to the formation of the bulge and/or disk. In elliptical galaxies, where the low-metallicity tail is weaker\footnote{This may be an observational selection effect due to the
large effective radii, or more extended metal-poor halo components of these galaxies.}, the metal-poor component may 
be the relic of the very earliest formation stage, whereas the metal-rich component originated in the former disks of the merged galaxies (e.g. \citet{harris07a}, \citet{bekki02}.

\subsubsection{The MDF of globular clusters} \label{sec:2.1.2}

The MDF of globular clusters in spiral galaxies and giant ellipticals is clearly bimodal
with the metal-rich clusters concentrated towards the galactic center and the metal-poor clusters
(which are invariably very old, $t \ge 10$ Gyr) 
distributed over larger distances \citep{gebhardt99,peng06,brodie12}. 
  For instance the MDF of GCs in the {\bf MW} has peaks at   
\FeH =$-1.55 \pm 0.07$ and $-0.55 \pm 0.10$ \citep{harris16} 
and the giant elliptical {\bf Cen A} has peaks at \FeH =-1.4 and -0.4 \citep{beasley08}.

The peak in the MDF of the metal-poor (blue) GCs was found to be 
almost constant at \FeH $\simeq$ -1.5 \citep{burgarella01} and showing no correlation 
with the parent galaxy luminosity \citep{forbes97} in several studies. However, more homogeneous
samples revealed a shallow but significant correlation between the peak MDF of the 
blue clusters and the host galaxy luminosity \citep[e.g.]{larsen01,strader04,peng06}. 
By contrast, 
the peak of the metal-rich (red) GCs depends strongly
on the mass or luminosity of the galaxies: 
the more massive a galaxy, the more metal-rich its red peak in the MDF 
(e.g.\  
\citet{peng06} for the {\bf Virgo galaxies}, and \citet{liu11} for the {\bf Fornax cluster} galaxies). 
The radial distribution of the metal-rich GCs in the {\bf MW} shows a steep decline beyond 
$R_{\rm Gal} \simeq 10$ kpc, whereas the population of metal-poor GCs extends all the way from 
about 1 to 100 kpc (e.g. see the review by \citet{harris01a}).
In {\bf Cen A} the majority of the clusters with \FeH $>$ -1 are within 20 kpc projected distance from the center,
whereas the metal-poor clusters show a flatter distribution with projected distance.
Also in large cD galaxies the galactic radial distribution of metal-rich clusters is much steeper than that of metal-poor clusters, see e.g. Fig. 20 of \citet{harris16} and \citet{harris17a}.

Several studies have shown that the radial distribution of the metal-rich GCs follow approximately 
the radial distribution of the star-light and that the kinematics of the these red clusters also follow that
of the stars, whereas the blue GCs have a larger velocity dispersion,  
e.g.  \citet{schuberth10, strader11, pota13} and the review by \citet{brodie06}.

\subsubsection{Giant Elliptical Galaxy NGC 5128 = Cen A}
\label{sec:cenA}

\begin{figure}
\centering
\includegraphics[width=11.0cm]{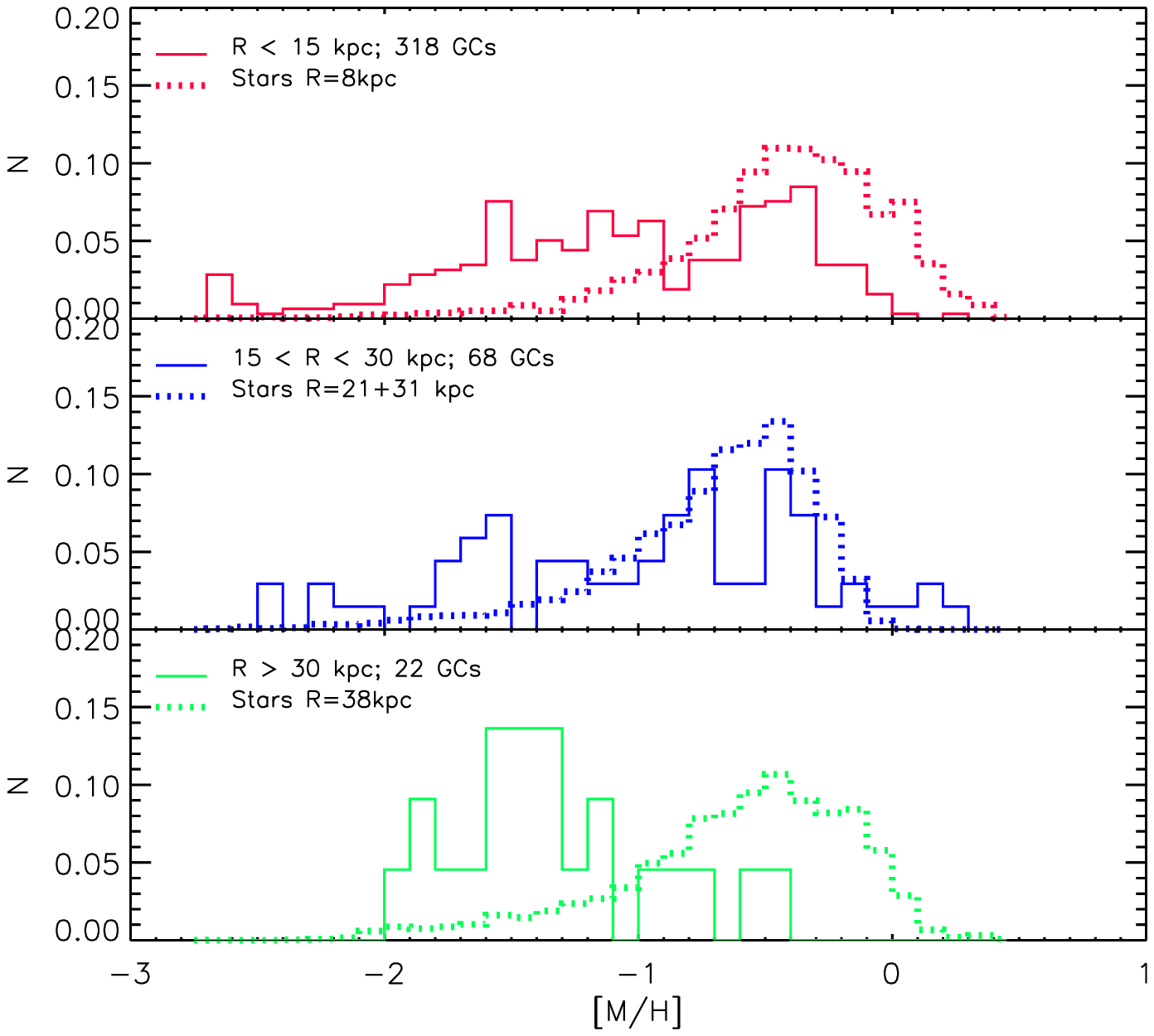}
\includegraphics[width=9.3cm]{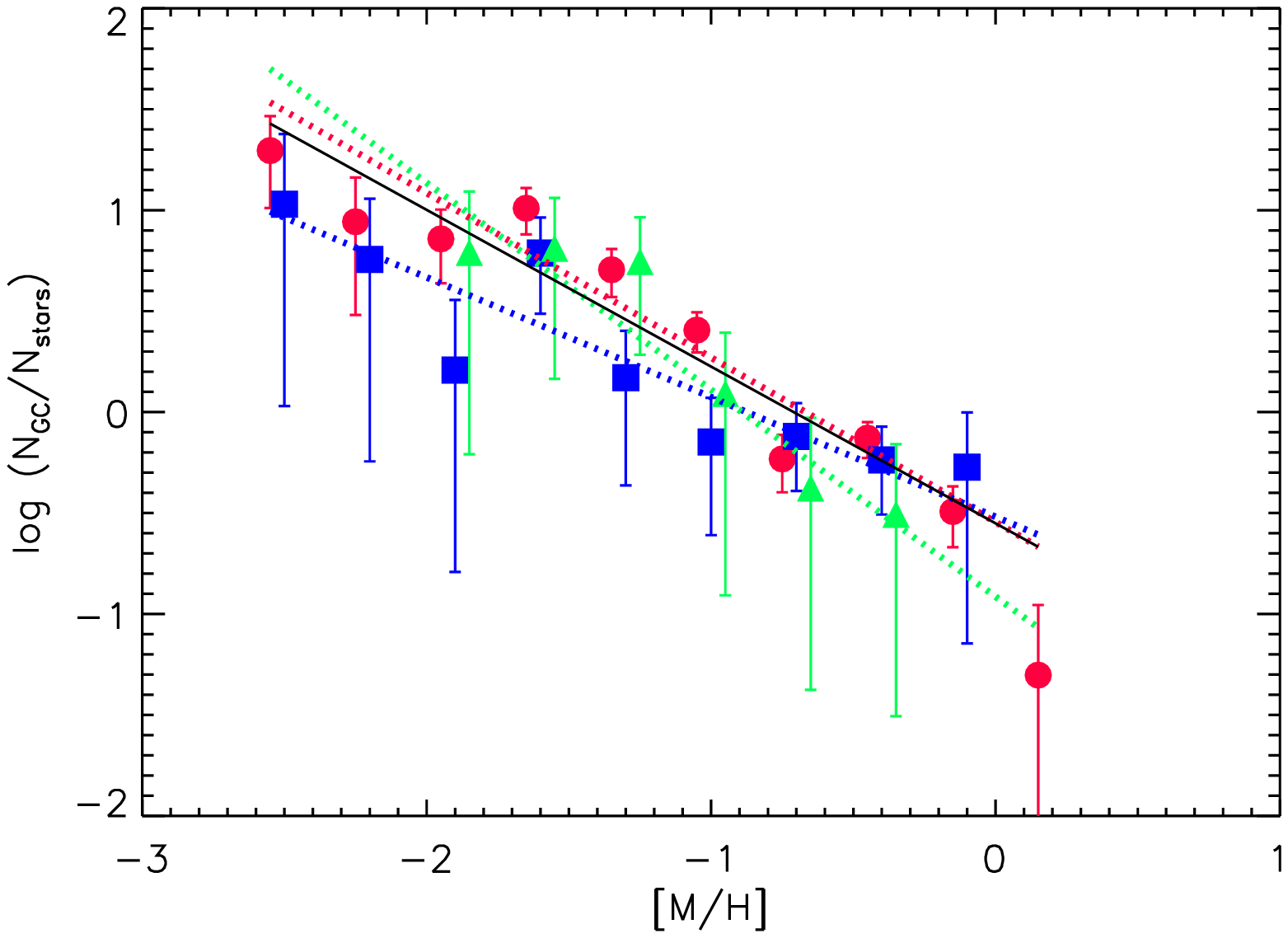}
\caption[] {Upper figure: Comparison between the normalized  MDF of stars (dotted) and GCs (full lines) 
in three distance regions of Cen A. Each MDF is normalized to 1.
Lower figure: The logarithmic ratio between the normalized MDFs of clusters and stars as a function of 
\MH\ in bins of $\Delta \MH = 0.30$. The colours are the same as in the upper figure. 
For clarity, the points of the three regions are slightly horizontal shifted with respect to one another.
The coloured dotted lines show the weighted least square fits of the three distance regions. The full
black line is the least square fit of the three regions combined. It has a slope of $-0.78 \pm 0.05$.
 }
\label{fig1}
\end{figure}

In Fig. \ref{fig1} the normalized MDF of GCs and halo stars in the giant elliptical galaxy 
Cen A (NGC 5218) at a distance of 3.8 Mpc \citep{harris10} are compared in three distance regions of the halo. 

The stellar MDF is based on resolved stellar population studies in fields at projected distances 
of (i) 8 kpc \citep{harrisharris02}, (ii) in two field at 21 and 31 kpc  \citep{harris99, harris00}, and (iii) 
in the outer halo field at 38 kpc galactocentric distance \citep{rejkuba05}. 
The MDF of the 408 GCs was derived by combining the samples of (i) 207 GCs with 
spectroscopic metallicities from \citet{beasley08} and (ii) 201 spectroscopically confirmed GCs with 
velocities consistent with being members of the globular cluster system (GCS) of NGC~5128, whose 
metallicities were calculated from  Washington (C--T1) photometric index \citep{woodley07}\footnote{See \citet{woodley10} for the  
calibration from (C--T1) colour to [Fe/H] metallicity.}. These GCs span a range of projected distances from 0 to 45 kpc, but 
the large majority is located within the inner 15~kpc -- this is primarily due to incomplete spectroscopic studies of
the outer regions of the GCS in this galaxy. 

The stellar MDF is measured by interpolating (V-I) colours of the upper Red Giant Branch (RGB) stars on a set of 
empirically calibrated stellar evolutionary tracks \citep{harris99} assuming an alpha enhancement
of [$\alpha$/Fe]=+0.3. The relative precision in MDFs of different fields is ensured by using the same
filters for observations of different fields, as well as the same colour-metallicity calibration
and interpolation code. While it is well known that the RGB colour varies non-linearly with metallicity 
and the photometric metallicities have higher errors at the metal-poor end \citep{saviane00,harris00}, 
comparison of photometric MDFs with those constructed from spectroscopic measurements of individual
stars shows a good agreement on the average values, as well as the overall shape
of MDFs \citep[e.g.][for a wide MDF in the Milky Way bulge]{gonzalez11}.\footnote{Note that \citet{gonzalez11} used 
near-IR photometric bands (J and K$_s$), and \citet{streich14} caution that metallicity determinations 
from RGB alone have an accuracy of 0.3 dex for simple stellar populations for [M/H]$\lesssim -1$ and 
0.15 dex for [M/H]$\gtrsim -1$.} \citet{kalirai06} also found a very good agreement between photometric and 
spectroscopic metallicities for M31 giants for  the whole metallicity range except the most metal-poor bin ($\FeH<-1.5$), where photometric measurements extend to lower values, making the photometric stellar MDF broader. The metallicity accuracy for GC 
measurements is of the order of $\sim$0.15 dex \citep{beasley08}, depending more strongly on the quality of spectra, 
then on the metallicity of individual clusters.

The top panel in the upper figure 
compares the MDF of the field stars at 8 kpc field with that of GCs located between $6 < R_p < 15$ kpc. 
The middle panel compares the MDF of GCs at $15 <R<30$ kpc with the stellar MDF in the combined 
fields at 21 and 31 kpc, which have very similar MDFs \citep{rejkuba05}.
The lower panel compares the MDFs of stars in the 38 kpc field with GCs at $R>30$ kpc.
We see that the stellar MDF in the inner region of Cen A is wider and extends to slightly higher $\MH$ than 
in the outer halo. \citet{harrisharris02} have argued that this is due to contributions by bulge stars. 
On the other hand, the MDFs of the GCs are flatter with the contribution of the metal-poor clusters 
with $[M/H] < -1.5$ increasing towards larger distances.

At all distances the clusters are underrepresented compared to stars at high
metallicity $\MH > -1$ and overrepresented at low metallicity, $\MH < -1$
(see also \citet{beasley08} for a similar observation). 

 (The terms ``underrepresented'' and ``overrepresented'' in this respect are always relative to the ``mean''
cluster-to-star ratio, which is generally of the order of a percent in mass).

The lower part of Fig. \ref{fig1} shows the logarithmic ratio between the 
{\it normalized MDFs} of clusters and stars
as a function of metallicity. In all three distance ranges this ratio decreases with increasing \MH\
by about 2 dex in the range of $-2.5~< [M/H]~< 0$.
\footnote{NB: This does not mean that the cluster to star ratio  $N_{\rm cl}/N_{\rm stars}$ is the 
same in the 3 regions because we used the normalized MDFs. The absolute values are not known, because 
the star-counts and the cluster-counts do not cover the same region of CenA.}
Although the absolute values of the specific frequency $S_N$ in the three distance regions is not known,
the trend in the lower part of figure \ref{fig1} shows that the decrease of $S_N$ with \MH\  
appears to be independent of distance. The weighted least square fits at the inner radial distance range 
has a slope of
$d \rm{log}(N_{\rm cl}/N_{\rm stars}) /d \MH ~= -0.82 \pm 0.06$. The slopes in the other two distance ranges 
are -0.59 and -1.0 but are less well determined due to the large error-bars. 
The linear regression of the full sample has a slope of $-0 78 \pm 0.05$. 
We conclude that the dependence of the
ratio $N_{\rm cl}/N_{\rm stars}$ on metallicity is independent of the galactocentric distance and has 
a slope of $d \rm{log}(N_{\rm cl}/N_{\rm stars}) /d \MH ~= -0.80 \pm 0.07$.
Further confidence in our conclusion is gained by comparing the stellar vs
GC MDFs in NGC~3115, which also show a similar dependence on metallicity that is independence of the 
distance, as shown in Fig.~11 and 12 of \citet{peacock15}.   

 \subsubsection{Spiral Galaxy M31} \label{sec:M31}

\begin{figure}
\centering
\includegraphics[width=12.0cm]{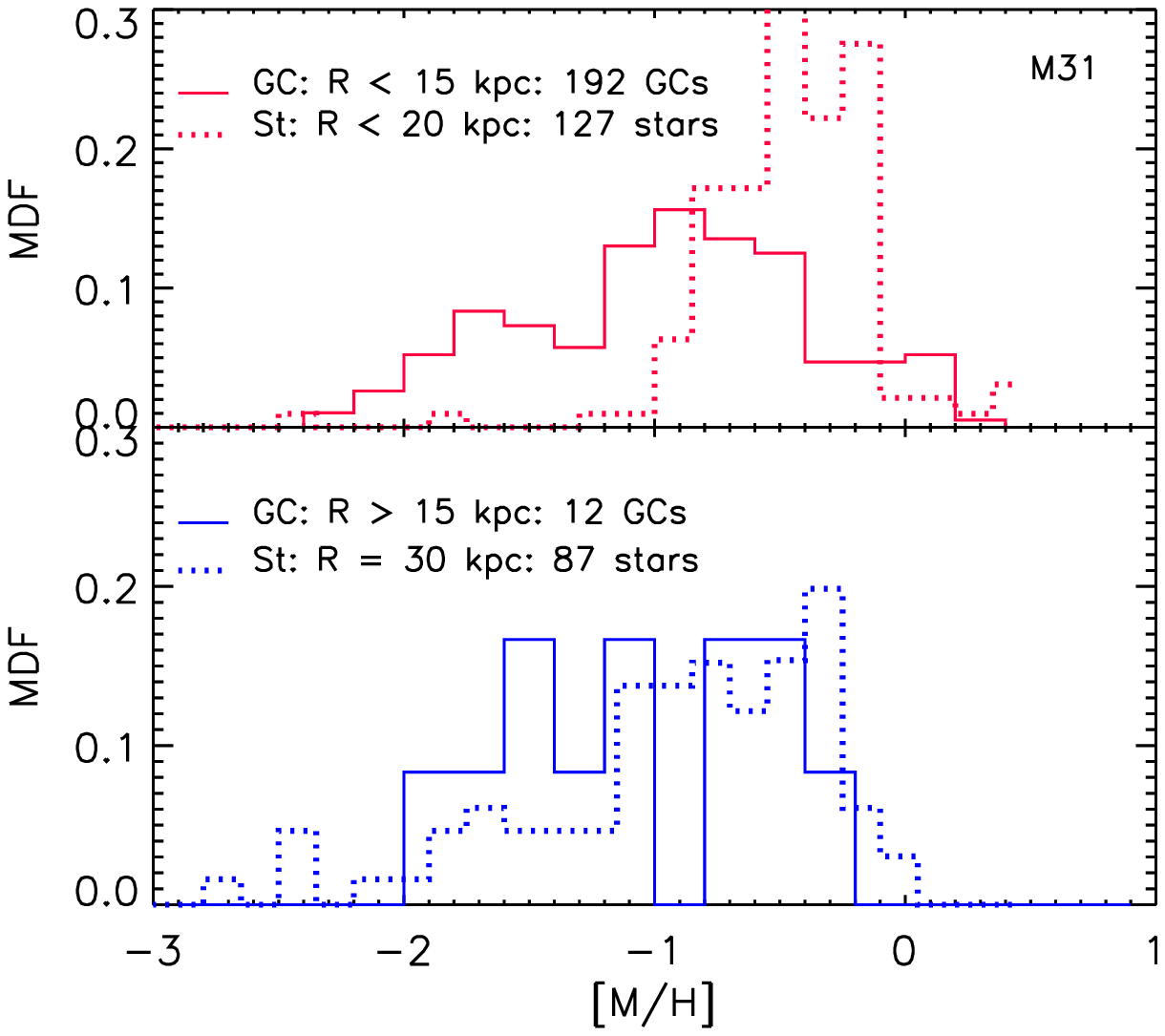} 
\includegraphics[width=9.0cm]{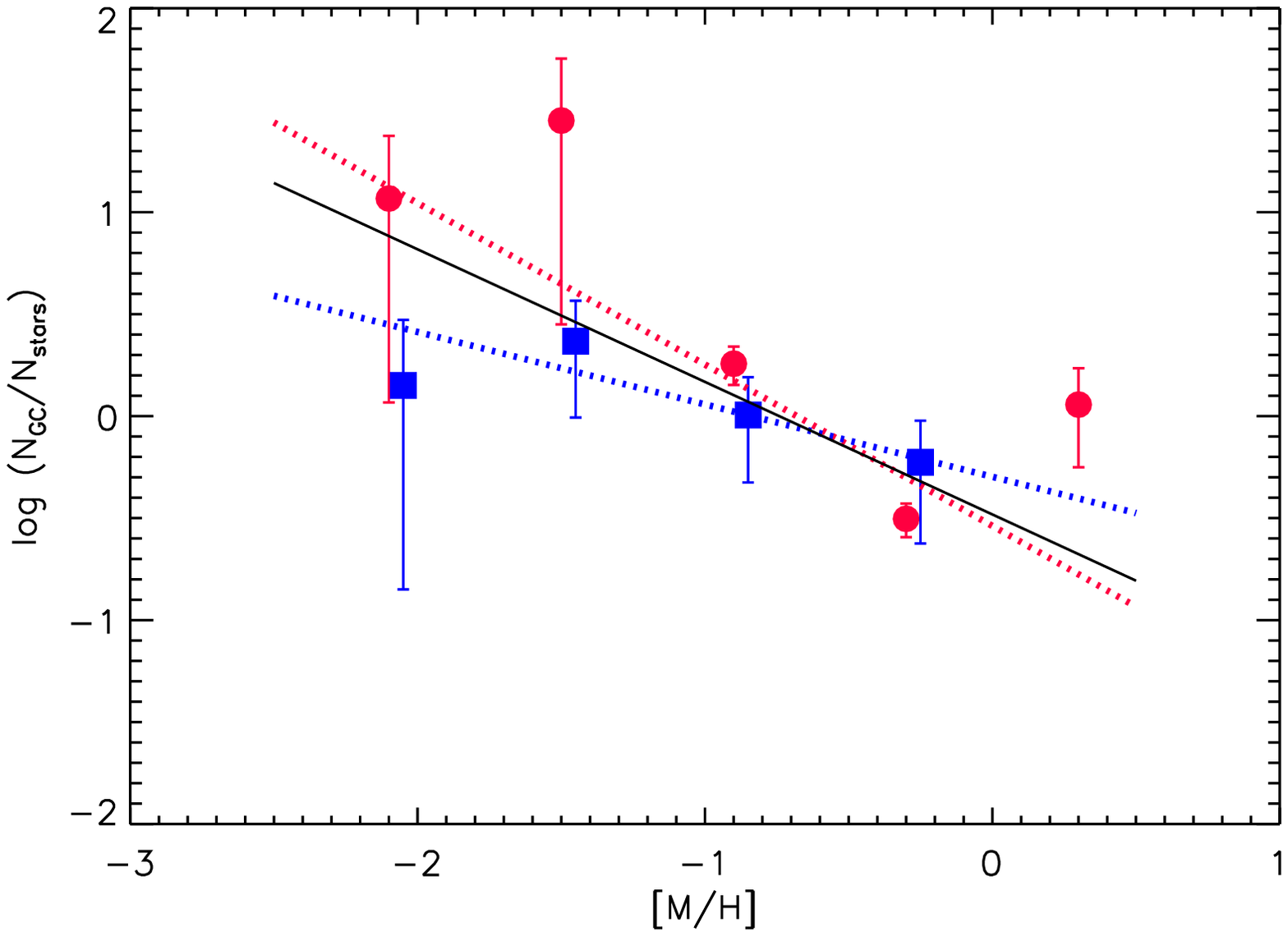} 
\caption[] {Upper figure: Comparison between the normalized  MDF of stars (dotted) and GCs (full lines)
in two distance regions of M31.
Each MDF is normalized to 1. Lower figure: The logarithmic ratio  between the normalized MDFs of clusters and stars as a function of [M/H] in bins of 0.6 at two distance regions of M31. 
The weighted least square fit of $(N_{\rm cl}/N_{\rm stars})$ versus \MH\ for the inner and outer regions 
are shown by dotted red and blue lines respectively. The black line is the fit to all data
and has a slope of $-0.69 \pm 0.27$.
}
\label{fig2}
\end{figure}

Fig. \ref{fig2} shows the normalized MDF of clusters and halo stars in M31.
The data for the clusters are taken from the Revised Bologna Catalog, the main
repository of information for M31 GCs  \citep{galleti04}, 
which contains metallicity measurements for 225 GCs based on Lick indices by \citet{galleti09}. 
The MDF of the halo stars is from \citet{kalirai06}. These authors derived the metallicities of 250
halo RGB stars in 12 fields, ranging in projected distance from 12 to 165 kpc from the center.
Assuming that the halo stars and the GCs are both distributed spherically around M31, 
rather than ellipsoidally,
we can directly compare the MDF of both samples as a function of projected galactocentric distance, 
by correcting for the tilt of the M31 disk.\footnote{Until more observations become available, this is presently the best one can do. \citet{ibata14} showed that the M31 most metal-poor populations ($\feh <-1.7$) are distributed approximately spherically with only a relatively small fraction residing in discernible stream-like structures. More metal-rich populations contain larger fractions of stars in streams with the stream fraction reaching 86\% at $\feh >-0.6$. So, actually the shape of the halo stellar distribution changes with metallicity.
It should be noted that also GCs are not homogeneously distributed and appear to be more abundant in the streams.}

The MDF of the halo stars is shown in two distance bins at $R<20$ and $R>20$ kpc. 
The GC sample was split at $R=15$ kpc to have sufficient clusters in the outer region to make a 
meaningful MDF. The inner region contains about 200 clusters, the outer region contains only 12 GCs

The figure shows that the MDF of the halo stars of M31 becomes wider and shifts to lower metallicity with
increasing distance (see \citet{kalirai06} for the gradual trend). In contrast, the MDF of the clusters
is approximately the same in both regions, within the statistical uncertainty. In both
the inner and the outer region the mean metallicity of the GC is lower
than that of the stars. The effect is strongest in the inner region which may, at least partly, be due 
the contribution by the bulge stars in the inner region \citep{kalirai06}. 

The lower part of figure \ref{fig2} shows the logarithmic ratio between the normalized MDF 
of clusters and stars as a function of \feh\ in bins of $\Delta \feh=0.6$.  
The linear weighted least square fit for the inner region has a slope of 
$d \rm{log}(N_{\rm cl}/N_{\rm stars}) /d \MH ~= -0.79 \pm 0.31$. 
The trend in the outer region is much less well defined, due to the fact that 
most bins contain only 1 to 4 clusters, so the uncertainty is large.
Combining all data we find a slope of $ -0.65 \pm 0.27$. This is compatible with the slope of $-0.80 \pm 0.07$ 
found for Cen A. (Fig. \ref{fig1}).\footnote{A similar 
comparison for the Milky Way is seriously hampered by the fact that
the stellar MDF can only be  measured at the solar radius,  $\Rgal(\odot)=8$ kpc, as a function of 
height $|Z|$ above or below the Galactic Plane where the  sample may be 
contaminated by thick disk stars up to $|Z| \sim 7$ kpc \citep{juric08}.}

\subsection{Dwarf galaxies}  \label{sec:obs:dwarfs}

\subsubsection{The MDF of stars in dwarf galaxies} \label{sec:2.2.1}

The stellar MDFs of the dwarf elliptical {\bf M32} \citep{grillmair96} and
irregular galaxy {\bf LMC} \citep{cole00, haschke12} are very similar to those of the
stellar halos of the giant elliptical Cen A and the spiral galaxy M31
\citep{harrisharris01}. They all show a peak around $\FeH \sim -0.5$
with a low metallicity tail extending to $\FeH \sim -2$.
Fainter dwarf galaxies ($-14<M_V<-10$), in particular dwarf elliptical (dEs) 
and dwarf spheroidals (dSphs),
show similar shapes of their stellar MDFs, but with much lower peak
metallicities of $\FeH ~ -1.5$ and extensions of their low metallicity tail
to $\FeH \sim -3$ ({\bf Local Group}: \citet{starkenburg10}; {\bf Cen\,A group}:
\citet{crnojevic10}; {\bf M81 group}: \citet{lianou10}).
The {\it mean metallicity} of the stellar populations of dwarf galaxies
shows a clear trend with galaxy mass and metallicity: the more massive or
luminous a galaxy, the higher its mean stellar metallicity, indicating that more
massive dwarfs had higher star formation rates or were better able to retain
the enriched gas that was released by stellar winds and supernovae \citep{grebel03}.
The same trend was also found for spirals and ellipticals.

Within dwarf galaxies, the metallicity may vary as a function of distance from the center.
For dwarf galaxies in the Local Group (LG), where metallicities can be
measured spectroscopically, the age-metallicity degeneracy problem in the interpretation of 
color-color distributions can be avoided.  
\citet{leaman13} derived the metallicity gradient in the LG dwarf irregular {\bf WLM} and compared it with 
those in the {\bf LG dSphs} and the {\bf Magellanic Clouds}. 
The metallicity 
gradient is small or absent in dIrrs, $d[M/H]/d(r/r_c) = -0.04 \pm 0.04$, but clearly present in dSphs,
$d[M/H]/d(r/r_c) \simeq -0.15 \pm 0.05$, where $r_c$ is the core radius, which is typically $~ 0.2$ to 0.6 kpc.    

\subsubsection{The MDF of globular clusters in dwarf galaxies}  \label{sec:2.2.2}

Dwarf galaxies have generally very small numbers of GCs so the best way to
derive information about their statistical properties is the combination of
results of many galaxies of similar types.
This has been done for samples of 69 dwarf ellipticals in the {\bf Virgo} and
{\bf Fornax clusters} \citep{miller07} and 57 and 68 nearby mainly dwarf irregulars  
by \citet{sharina05} and \citet{georgiev09}.

Globular clusters in dwarf galaxies are in general very metal-poor, mostly even more
metal-poor than the peak metallicity of the dwarf's stellar MDF
(e.g. \citet{mackey04, sharina10}). 
The {\bf Fornax dSph}, for example,
contains four GCs with $\FeH \sim-2.5$ to -1.9 and one more metal-rich GC
at $\FeH \sim-1.5$ dex, whereas its stellar MDF peaks
at $\FeH \sim-1.0$ with a long tail towards lower metallicities \citep{strader03, helmi06, larsen12b}. 
A similar situation is found in the dwarf galaxy {\bf WLM}, where the GCs
with a metallicity of  $\FeH =-2.0$ are significantly more metal-poor than the
average of -1.3 of the field stars in this galaxy \citep{leaman13, larsen12}. The dwarf galaxy {\bf IKN} has five GCs but only the metallicity $\FeH =-2.1$ of IKN-5 is known with some confidence. The average metallicity of the field stars is between -1.5 an -1.0. \citep{lianou10, georgiev10, larsen14}. 

\begin{figure}
\vspace{-3.5cm}
\includegraphics[width=12.0cm]{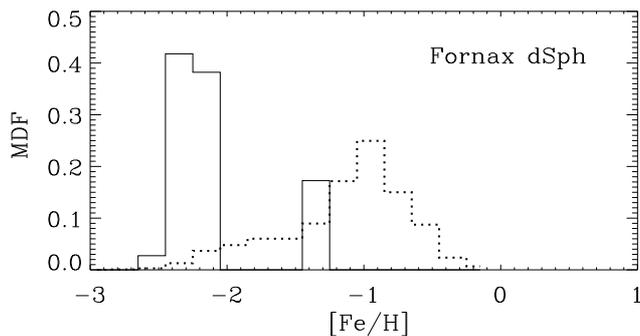} \vspace{-1.0cm} 
\caption[] {Comparison between the normalized  MDF of stars (dotted) and GCs (full lines) in
the Fornax dwarf spheroidal galaxy. Each MDF is normalized to 1.}
\label{fig3}
\end{figure}

Fig. \ref{fig3} shows the MDF of stars and GCs in the Fornax dSph, 
based on the data by \citet{larsen12b}. It shows the same trend as the MDFs in the outer halo of CenA but
more extreme: the MDF of the stars peaks near $\feh \simeq -1$ with a low metallicity  tail to $\simeq$ -2,
but the MDF of the GCs is narrower and at lower metallicity than in CenA. 
The GCs at $\feh > -1$ are conspicuously absent. 

\subsection{The specific frequency of clusters in different galaxies} \label{sec:2.3}

The ratio between the number of clusters and the stellar population can be expressed in terms of the
specific frequency, $\Sn = N_{\rm GC} \times 10^{0.4(M_V+15)}$ which is the number of GCs relative to the B-magnitude of the Galaxy \citep{harris81} or the specific luminosity $T_L=100\times L_{\rm GC}/L_{\rm Gal}$ 
\citep{harris91} or the specific mass $S_m=100 \times M_{\rm GC}/M_{\rm Gal}$, where 
$M_{\rm Gal}$ is the stellar mass 
\citep{zepf93, peng08}. The dependence of $\Sn$ or $T$ on galaxy type has been reviewed by \citet{brodie06}. 
The largest dataset of about 100 early type 
galaxies in the {\bf Virgo cluster}, shows that the specific mass $S_m$ has a typical low value of about 0.3
for galaxies in the mass range of $9.5 < \log (M_{\rm Gal}/\Msun) < 11.0$ \citep{peng08}.  
This implies that in these galaxies the mass fraction of 0.003 is in clusters. 
Beyond this mass range, the mean values of $S_m$ increase to both lower and higher mass. Galaxies with 
$M_{\rm Gal} \simeq 10^{12}$ \Msun\ reach $S_m \sim 1$. Galaxies with  $M_{\rm Gal} < 10^9$ \Msun\ 
have a large scatter in $S_m$ between 0 and about 2, with 
nearly all dwarfs with large $S_m$ located within 1 kpc form the cD galaxy {\bf M87}. This
suggests that the cluster formation in dwarfs is biased towards dense environments.

An observable that is key to the present work is the relative specific frequency of 
metal-poor and metal-rich GCs as a function of galaxy mass. In \citet{peng08}, $S_N$ is derived 
for both subpopulations. 
In the mass range of $10^9$ to $10^{11}$ \Msun\ the value of $T$, which is the total 
number of clusters per $10^{9}$ \Msun\ 
 decreases with galaxy mass from about 30 at $M=10^9$ \Msun\ to 0.5 at $M=10^{11}$ \Msun.
Over this mass range the fraction of the blue (metal-poor) clusters is between 90 and 70 \%.
 Since the metallicity of these galaxies increases with increasing mass, it indicates a
decrease of $S_N$ with increasing metallicity. This trend is similar to the one found in {\bf Cen A} (Fig.1).
In galaxies with $M>10^{11}$ \Msun\ 
the metal-rich fraction flattens or even decreases somewhat, which is likely the result of 
a substantial fraction of the GC populations in these galaxies stemming from accreted 
dwarf galaxies.

In galaxies for which the radial distributions of the stellar light and the GC systems both have been measured,
the stellar light overall decreases faster with radius than the GC systems. This indicates that the specific frequency or the value of
$T$ of the GCs {\it increases} with distance. If the GC system is split into red 
metal-rich clusters and blue metal-poor clusters, the red component always drops steeper with distance and 
follows the stellar light profile more closely than the blue component. This indicates that $T$ of the blue 
component increases faster with distance than the red component \citep{brodie06, pota13}. This
is in comparison to total stellar light. However, \citet{peacock15} have shown for {\bf NGC~3115} that if the 
samples of GCs and stars are both split into blue and red (metal-poor and metal-rich)
components, the blue GCs follow the shallow slope of the metal-poor stars.
So, the metal-poor clusters follow the same radial distribution as the metal-poor stars and 
the metal-rich clusters follow the radial trend of the metal-rich stars. Consequently, the cluster-to-star 
ratio is independent of distance and only dependent on metallicity (in agreement with Figs. 1 and 2 for 
{\bf Cen A} 
and {\bf M31} respectively.)


\section{Summary of  the observations} \label{sec:obssum}

\begin{table*} \label{tbl:summary}
\caption[]{The contribution of high and low-metallicity objects to the normalized MDFs of stars and clusters 
in the inner and outer halos of galaxies.}
\begin{tabular}{l l l | l l| l l }
\hline
\hline
Galaxy  &  Boundary &  Objects & Inner  & Inner &  Outer & Outer \\
        &           & & $\feh<-1$ & $\feh>-1$ &  $\feh<-1$ & $\feh>-1$ \\
\hline
Cen A   &  15 kpc   & Stars    & 0.095  & 0.905 & 0.159 & 0.841 \\
        &           & Clusters & 0.330  & 0.670 & 0.522 & 0.478 \\
        &           & log(\Rgcst ) & 0.54 & -0.13 & 0.52 & -0.24 \\
\hline
M31     &  $\sim$ 17 kpc &  Stars    & 0.029  & 0.971 &  0.351 & 0.648 \\
        &                &  Clusters & 0.432  & 0.568 &  0.643 & 0.357 \\
        &                &  log(\Rgcst) & 1.17  & -0.23 & 0.26  & -0.26 \\
\hline
Fornax dSph &             & Stars  & 0.482 & 0.517 & & \\
           &             & Clusters  & 1.00 & 0.00 & & \\
           &             & log(\Rgcst) & 0.32 & $<$-3 & & \\
\hline
\hline\\  
\end{tabular}

(a) Boundary = separation between inner and outer region. (b) The sum within each region is 1. 
(c) \Rgcst\ is the ratio between the contribution by clusters of a certain metallicity range to the 
normalized cluster MDF and the contribution by stars in the same metallicity range to the stellar MDF.

\end{table*}

Table \ref{tbl:summary} summarizes the differences in the contributions of metal-poor, $\Feh < -1$, and metal-rich,
$\Feh > -1$, stars and clusters at the inner and outer regions of the large galaxies Cen A and M31 and in the Fornax dSph.
For each group of objects (stars or clusters) the sum of the contributions of the high and low metallicity
fraction is normalized to 1 in each region.
The table also shows the ratio of the cluster-vs-stellar contributions to the MDF at the
inner and outer regions of these galaxies, with the adopted boundary indicated. Notice the general trends that:\\ 
(a) in both the inner and outer regions the relative contribution of the clusters at low metallicity is 
larger than that of the stars, \\
(b) in both the inner and the outer regions the relative contributions of the clusters at high metallicity is 
smaller than that of the stars, \\
(c) the ratios of the Fornax dSph are more similar to those in the
outer regions of the larger galaxies than to the inner regions, with the note that there is not a single 
cluster at $\feh > -1$. (Due to the low number statistics of only 5 clusters in this galaxy, 
the value of log(\Rgcst) would have increased to 
-0.5 if only 1 metal-rich cluster had been found in the Fornax dSph.)\\ 
(d) In the inner and outer regions of the halo of the large spiral galaxy M31 the cluster-to-star ratio is high
at low metallicity and low at high metallicity.\\
(d) In dwarf galaxies both the stellar populations and the GC populations are metal-poor, but the clusters are more metal-poor than the stars.

The observations summarized above raise the following questions:
\begin{enumerate}
\item[{\bf Q1.}] Why are the GCs in dwarf galaxies on average more metal-poor than the stars?
\item[{\bf Q2.}] In the inner regions of the halos of large galaxies that are dominated by metal-rich stars, 
why are there no or very few metal-rich clusters (left)? 
\item[{\bf Q3.}] In the outer regions of the halos of large galaxies, that are dominated by metal-poor clusters, 
why is the metal-poor stellar population that should accompany the clusters absent or very weak? 
\item[{\bf Q4.}] Why does the cluster-to-star ratio, expressed as \Rgcst, decrease so strongly with increasing metallicity in giant elliptical (CenA), in spiral galaxies (M31) and dwarf ellipticals 
(Fornax dSph).? 
\item[{\bf Q5.}] Why is this trend approximately independent of projected radial distance in CenA and M31?
\end{enumerate}

To answer these questions we have to consider the formation of clusters in different environments
as well as their destruction by various effects. In \S 4 we first discuss the formation and in 
\S 5 and \S 6 we will discuss the destruction of GCs.


\section{The formation of globular clusters} \label{sec:form}

In \S\ref{sec:form}--\ref{sec:cosmology} we address possible explanations 
for the observed differences
 between the MDFs of halo stars and globular clusters. In order for any explanation to be viable, it should either depend on the metallicity itself, or have an indirect dependence on metallicity due to differences in cosmological environment. We do not consider the metallicity dependence of the star formation rate, because this study concentrates on the relation between clusters and stars which depends only on the cluster formation efficiency (CFE), i.e. the fraction of star formation that ends up in bound clusters, and on cluster destruction. Because more massive clusters have a higher survival probability (see \S\ref{sec:destruc}), any environmental dependence of the maximum cluster mass could also be important in setting the globular cluster-to-field star ratio. In this section, we first focus on the mechanisms related to globular cluster {\it formation} and any resulting environmental variation of the CFE, residual gas expulsion, and the maximum cluster mass.
 
Throughout \S\ref{sec:form}--\ref{sec:cosmology}, the presented calculations and arguments have been motivated by the observed physical conditions of the ISM and star formation in galaxies at high redshift ($z=1$--$3$), obtained either through adaptive optics observations, sub-mm interferometry, or gravitational lensing \citep[e.g.][]{forsterschreiber09,genzel11,swinbank11,swinbank12,livermore15,dessauges17}. In (almost) all cases, the environmental dependences of the considered physical mechanisms have been identified and applied in this high-redshift context.

\subsection{The cluster formation efficiency and its dependence on metallicity} \label{sec:form:cfe}

The zeroth-order solution to the metallicity-dependence of the cluster-to-field-star mass ratio is that the fraction of star formation that results in bound stellar clusters (the cluster formation efficiency or CFE $\Gamma$, see \citealt{bastian08}) is a function of metallicity. Specifically, the CFE should decrease with metallicity in order to simultaneously produce the observed MDFs of clusters and field stars. It is therefore relevant to address whether (and how) the CFE can be affected by metallicity or the environment. 

Globular clusters are thought to have formed in high-pressure, actively star-forming environments \citep[e.g.,][]{ashman92,elmegreen97,elmegreen10,shapiro10,kruijssen15b}. The observed metallicity dependence of the cluster-to-star ratio could be explained if the CFE in metal-rich star-forming galaxies is lower than in metal-poor star-forming galaxies. There is a broad relation of increasing metallicity with galaxy mass, from low to high redshift \citep[e.g.][]{tremonti04,erb06}, which can be understood by arguing that metals cannot be retained in the shallow potential wells of low-mass galaxies \citep[e.g.][]{maclow99}. 
This clear and systematic trend of increasing metallicity with galaxy mass has been observed out to $z\sim4$ down to stellar masses of $M_\star\sim5\times10^8~\msun$ and metallicities ${\rm [Fe/H]}\sim-1.3$, with a dependence of decreasing metallicity with redshift \citep{mannucci09}.
 For this reason, metal-poor globular clusters have been proposed to originate from dwarf galaxies \citep{elmegreen97,muratov10}, whereas metal-rich globular clusters were likely produced in more massive galaxies \citep{forbes97,shapiro10,kruijssen12c}. 
This is a reasonable argument, even if the present-day globular cluster population of a galaxy spans a broad metallicity range, because galaxies are widely thought to have formed hierarchically from a large number of lower-mass progenitors.

The CFE increases with the gas surface density of the galaxy disc $\Sigma_{\rm g}$, because higher-pressure gas can achieve higher star formation efficiencies, resulting in a larger fraction of all star formation in bound clusters \citep{kruijssen12d}. In the local Universe, the gas pressure does not exhibit a strong dependence on galaxy mass, as high gas or star formation rate surface densities are observed across the galaxy mass range \citep[see e.g.~the compilation by][]{adamo15b}. However, \citet[Table~1]{kruijssen15b} propose that despite considerable (and possibly dominant) scatter, the higher binding energy of massive galaxies leads to a weak increase of the gas pressure with galaxy mass and (hence) metallicity at a given redshift. 
The predicted CFEs for globular clusters thus increase with metallicity, which for the observed galaxy mass-metallicity relation at $z\sim3$ \citep{mannucci09} leads to $\Gamma=0.38$ at ${\rm [Fe/H]}=-2.3$ (this requires an extrapolation of the mass-metallicity relation to metallicities ${\rm [Fe/H]}<-1.3$) and $\Gamma=0.56$ at ${\rm [Fe/H]}=-0.6$ \citep{kruijssen15b}.
 This result cannot explain the observations mentioned in \S\ref{sec:obs}, because they require a higher cluster formation efficiency in low metallicity galaxies.\footnote{
In addition, we note that the {\it observed full dynamic range of the CFE} is only a factor of 50 \citep[e.g.][]{goddard10,adamo15b,johnson16}. Even if one would extend this to the extreme case of $\Gamma=1$, the total dynamic range of the CFE is two orders of magnitude. For the conditions under which GCs are likely to form ($P/k=10^6$--$10^9~{\rm K}~{\rm cm}^{-3}$, \citealt{elmegreen97}), the CFE is only expected to vary by a factor of $\sim2$ \citep[$\Gamma=0.3$--$0.6$,][]{kruijssen12d}. This range is considerably smaller than the required dynamic range of three orders of magnitude in the star-to-cluster ratios across the metallicity range (see \S\ref{sec:obs}).}

In principle, one could also argue that the redshift dependence of galaxy metallicity signifies higher formation redshifts for metal-poor globular clusters than metal-rich ones. This is only partially true observationally \citep[e.g.][]{forbes10}, but even then the dynamic range of metallicity is too limited for this idea to work. For a Milky Way progenitor galaxy of $M_\star\sim10^9~\msun$ at $z=4$ \citep[cf.][]{moster13}, the metallicity is ${\rm [Fe/H]}\sim-1$. If the metallicity would only depend on redshift (rather than on galaxy mass during hierarchical galaxy formation), the metal-poor globular clusters should have formed at $z\gg4$, where gas pressures and CFEs were likely lower than at the peak of the cosmic star formation history \citep[$z\sim2$, e.g.][]{madau14}, leading to a peak CFE at ${\rm [Fe/H]}>-1$. This is again in disagreement with the trend of monotonically increasing cluster-to-star ratios towards low metallicities. Metallicity is therefore primarily a tracer of galaxy mass, and only secondarily of redshift.

{\it These theoretical arguments imply that it is unlikely that a metallicity dependence of the CFE can explain the differences between the MDF of halo stars and globular clusters, because it is opposite to the observed metallicity trend of the cluster-to-star ratios. Additionally, the dynamic range of the observed cluster-to-star ratios across the range of GC metallicities in galaxies (spanning three orders of magnitude) is unlikely to be explained by differences in the CFE as that is only expected to vary by a factor of $\sim2$.}

\subsection{The destruction of clusters at young age by gas expulsion (infant mortality)} \label{sec:form:expulsion}

It has been proposed that, after the formation of a bound stellar cluster is completed, its long-term survival is not necessarily guaranteed due to residual gas expulsion \citep[e.g.][]{lada84,boily03,baumgardt07}. The cluster could end up unbound at an early phase if the star formation process was locally inefficient, so that a significant fraction of the cluster mass is in the form of gas. The removal of this gas by stellar winds and radiative heating may unbind the cluster. This will depend on the gas fraction 
that is expected to remain when feedback obstructs further star formation.

Recent observational and theoretical evidence shows that the star formation process can lead to initially virialised and gas-poor stellar systems before any significant feedback can have taken place \citep{kruijssen12,girichidis12b,cottaar12,longmore14}. This implies that newly formed massive clusters do not contain a large amount of gas and hence that the influence of gas expulsion (infant mortality) on the destruction of massive clusters is small \citep[cf.][]{longmore14}.\footnote{The expulsion of residual gas may lead to the dispersal of young stellar associations, but they are irrelevant to the question at hand because we are considering massive clusters.}

{\it Based on the current theoretical picture of cluster formation, infant mortality by gas expulsion is unlikely to explain the observed differences between the MDF of globular clusters and halo stars, provided that globular cluster formation proceeds in a similar way to massive cluster formation observed in the local Universe.}

\subsection{The environmental dependence of the maximum cluster mass} \label{sec:form:mmax}

As we will discuss in \S\ref{sec:destruc}, the disruption timescale of clusters in a given environment increases with the cluster mass. This means that the decreasing cluster-to-star ratio with metallicity could be explained if massive clusters are less likely to form in high-metallicity environments. This possibility is explored here.

There is a range of observational evidence showing that the cluster mass function has an exponential truncation at the high-mass end, both for young clusters \citep[e.g.][]{gieles06a,larsen09,portegieszwart10,johnson17} and globular clusters \citep[e.g.][]{fall01,jordan07,kruijssen09b}. At the same time, it is clear that this maximum mass scale is not universal. Across the aforementioned range of papers, the maximum cluster mass covers $M_{\rm c}=8\times10^3$-- $3\times10^6~\msun$. It is an important question what causes this large dynamic range and how it may depend on the metallicity.

It was suggested by \citet{kruijssen14c} that the maximum cluster mass follows from the maximum giant molecular cloud (GMC) mass as $M_{\rm c}=\epsilon_{\rm sf}\Gamma M_{\rm GMC}$, with $\epsilon_{\rm sf}$ the star formation efficiency (SFE) and $M_{\rm GMC}$ the maximum GMC mass, which is proposed to be set by the \citet{toomre64} mass, i.e.~the largest self-gravitating scale in a differentially rotating disc. This hypothesis is supported by the good agreement with observed maximum GMC and cluster masses across different galactic environments \citep{kruijssen14c,adamo15b,freeman17}. \citet{reinacampos17} generalized this model by showing that in environments with low gas surface density and low angular velocity (i.e.~shear), the maximum cluster mass is not limited by shear, but by feedback disrupting a collapsing cloud before it has coalesced altogether. This updated model yields excellent agreement with observations of clouds, clumps, and clusters from the local Universe out to high redshift, including the lowest maximum cluster mass observed to date \citep[$M_{\rm c}=8\times10^3~\msun$ in M31,][]{johnson17}. Most importantly for the present work, \citet{reinacampos17} demonstrated that the high-pressure conditions of globular cluster formation (almost) entirely correspond to the shear-limited regime.

In the context of globular cluster formation, \citet{jordan07} observed that globular cluster mass functions have maximum mass scales $M_{\rm c}$ covering a dynamic range of roughly one order of magnitude ($M_{\rm c}=4\times10^5$--$4\times10^6~\msun$) in Virgo cluster galaxies, in such a way that $M_{\rm c}$ increases with galaxy mass $M_\star$. \citet{kruijssen15b} combined the $M_{\rm c}$--$M_\star$ relation with the relation between the globular cluster system metallicity and galaxy mass from \citet{peng06} to derive an approximate relation of $\log{(M_{\rm c}/\msun)}\sim6.5+0.7{\rm [Fe/H]}$ and subsequently use this relation to constrain the ISM conditions as a function of metallicity assuming that the maximum globular cluster mass is indeed shear-limited. As a result, it is found that gas pressures, maximum GMC masses, and (as is shown by the above expression) maximum cluster masses all increase with metallicity, analogously to the CFE in \S\ref{sec:form:cfe}. This is opposite to the required trend.

While the metallicity dependence of $M_{\rm c}$ has the wrong sign to explain the observed cluster-to-star ratios, the maximum cluster mass likely plays at least some role in setting the extremes of the globular cluster MDF relative to that of field stars. Because $M_{\rm c}$ is thought to decrease with cosmic time due to a decrease of the gas pressure \citep{reinacampos17}, fewer massive (and thus long-lived) clusters per unit field star mass are formed at later times. Due to the increase of the metallicity with cosmic time, this means that the cluster-to-star ratio at high metallicities is lowered further by continued star formation after the peak epoch of globular cluster formation has passed. However, these young stars will not be co-spatial with the older stars and clusters due to differences in the scale height and scale radius of the star-forming disc and the old stellar population. Massive galaxies formed through many minor and/or major mergers and have rich haloes. We therefore do not expect continued star formation to affect the cluster-to-star ratios that we obtain for the galaxy {\it haloes} of Cen A and M31 in several radial bins. It is most likely that the extension of the field star MDF to higher metallicities than globular clusters is simply in accordance with the monotonic (i.e.~featureless) decrease of the cluster-to-star ratio towards high metallicities in these galaxies (see \S\ref{sec:obs}). By contrast, continued star formation may have contributed to the low cluster-to-star ratio at high metallicities seen in Fornax, because we cannot separate out the halo due to the galaxy's dSph morphology.

The MDF of field stars does not only extend further than that of globular clusters at high metallicities, but it also extends further at the low-metallicity end -- there are no known globular clusters with metallicities ${\rm [Fe/H]}<-3$ (see \S\ref{sec:obs}), but extremely metal-poor field stars do exist \citep[e.g.][]{frebel15}. Because chemical enrichment proceeds extremely rapidly at low metallicity, such stars must have formed at low star formation rate surface densities and thus gas pressures, inhibiting the formation of massive and long-lived clusters. We therefore suggest that the apparent minimum globular cluster metallicity results from a strong decrease of both $M_{\rm c}$ and $\Gamma$ at ${\rm [Fe/H]}\lesssim-2.5$.

{\it The available observational and theoretical evidence suggests that the maximum cluster mass increases with metallicity at high redshift and therefore cannot explain the dearth of globular clusters at high metallicities relative to field stars, which would require the maximum cluster mass to decrease with metallicity so that fewer long-lived clusters are formed towards high metallicities.}


\section{The destruction of bound globular clusters} \label{sec:destruc}

Bound clusters that survived the early destruction processes mentioned above, may later dissolve by 
the loss of stars due to stellar evolution, two-body relaxation with tidal evaporation, and tidal shocks.
This dissolution is an important 
ingredient for understanding the relation between the populations of stars and clusters, because 
it reduces the cluster population. Cluster dissolution also contributes to the field star population,
but this effect is negligible because only a very small fraction of stars is born in  
bound massive clusters.

The destruction of star clusters is expected to depend on (their location in) the host galaxy. 
Because the metallicity is a strong function of galaxy mass and since most galaxies have a radial metallicity gradient, with metallicity increasing towards 
the center, any environmentally-dependent cluster destruction will result in a metallicity-dependent 
cluster destruction. In principle this might explain qualitatively why
the metal-poor and metal-rich globular cluster populations follow different spatial distributions 
within the halo of their host galaxy, with metal-rich clusters predominantly residing at small 
galactocentric distances ($ R < 5$ kpc in the Milky Way, \citealt{harris96}).

\subsection{Cluster destruction by stellar evolution and tidal evaporation} \label{sec:stripping}

Clusters orbiting in a galaxy lose stars by tidal evaporation. The mass evolution due to tidal evaporation 
and stellar evolution can be described by 

\begin{equation}
\frac{M(t)}{M_i}\simeq 
\left\{(\mu_{\rm ev}(t))^{\gamma}-
\frac{\gamma t}{t_0^{\rm tidal}}\left(\frac{\Msun}{\Mi}\right)^{\gamma}
\right\}^{1/\gamma}
\label{eq:muapprox}
\end{equation}
with masses in units of \Msun\ and  $\gamma \simeq 0.65$ \citep{baumgardt03,lamers05,lamers10}.
In this expression $\mu_{\rm ev}(t)$ is the fraction of the mass that is lost by stellar evolution, with
$\mu_{\rm ev}(t) \simeq 0.3$ at $t> 10$ Gyr. Eq. \ref{eq:muapprox} shows that clusters with an {\it initial mass} of $\Mi\simeq 10^{6}$ \Msun\
will be destroyed by tidal effect on a timescale of $t \simeq 9 \times 10^3~t_0^{\rm tidal}$.
The dissolution parameter $t_0^{\rm tidal}$ for clusters orbiting the center of a galaxy with a logarithmic
potential (i.e. a flat rotation curve $V_{\rm Gal}$) is

\begin{equation}
t_0^{\rm tidal} = 2.6~  R_{\rm Gal}~ (V_{\rm Gal}/220~\kms)^{-1} ~ ( 1-\epsilon)~~ {\rm Myr}
\label{eq:tzero}
\end{equation}
where $R$ (in kpc) is the apocenter and $\epsilon$ the eccentricity of the cluster orbit
\citep{baumgardt03}. This results in $t_0^{\rm tidal} \simeq 22$ Myr for Milky Way clusters on circular orbits at 8.5 kpc. Clusters with $M_i = ~1~10^6~\Msun$ will be destroyed within 12 Gyr if $t_0^{\rm tidal} < 1.3$ Myr.  Clusters on circular orbits in spiral galaxies ($V_{\rm Gal}\simeq 250$ \kms ) will only be destroyed by tidal destruction if they are at $R_{\rm Gal} < 0.5$ kpc and those in giant elliptical galaxies ($V_{\rm Gal} \simeq 500$ \kms) at $R_{\rm Gal} < 1.2$ kpc. 

{\it We conclude that 
tidal evaporation and evolution alone can only explain the small metal rich cluster-to-star ratio 
observed in Cen A and spiral galaxies near their very center. However, the observed trend extends 
over a much wider distance range.}

\subsection{Cluster destruction by dynamical friction} \label{sec:dynfric}

In the previous section we have assumed that the orbits of clusters do not change during their lifetime.
We now consider the decrease of cluster orbits due to dynamical friction. 

Massive objects that are moving in a dense medium consisting of lower-mass bodies experience a gravitational drag that is caused by the gravitational attraction of these bodies towards the wake behind the massive object. Under the right conditions, this dynamical friction  can cause massive globular clusters to spiral in towards the centers of their host galaxies \citep[e.g.][]{chandrasekhar43,tremaine75,capriotti96,lotz01}, where they may be destroyed by tidal evaporation. A simplified expression for the timescale on which an object spirals into the galactic center due to dynamical friction is given in \citet{binney87} under the assumption of a flat galaxy rotation curve:
\begin{equation}
\label{eq:tdf}
t_{\rm df}=\frac{264~{\rm Gyr}}{\ln{\Lambda}}\left(\frac{R_{\rm i}}{2~{\rm kpc}}\right)^2\left(\frac{\Vgal}{250~{\rm km}~{\rm s}^{-1}}\right)\left(\frac{M}{10^6~\msun}\right)^{-1} ,
\end{equation}
where ln $\Lambda\approx 10$ is the Coulomb logarithm of the encounters, $R_{\rm i}$ is the initial galactocentric radius, $\Vgal$ is the circular velocity of the host galaxy, and $M$ is the cluster mass. As the formula shows, dynamical friction is more efficient (i.e. the dynamical friction timescale is shorter) for small initial radii, small circular velocities, and large cluster masses. This latter property of dynamical friction contrasts with the dynamical disruption that is discussed in Sect.~\ref{sec:stripping}, which favors the destruction of low-mass clusters.

Assuming a mean globular cluster mass of $5\times10^5~\msun$ over the course of their lifetime to account for stellar evolution and dynamical mass loss, Eq.~\ref{eq:tdf} provides a critical initial galactocentric radius within which any globular clusters of age $t$ must have spiraled-in due to dynamical friction, as a function of the circular velocity:
\begin{equation}
\label{eq:rcrit}
R_{\rm crit}\approx \left(\frac{t}{12~{\rm Gyr}}\right)^{1/2}\left(\frac{\Vgal}{250~{\rm km}~{\rm s}^{-1}}\right)^{-1/2}~{\rm kpc} .
\end{equation}
For our Galaxy ($\Vgal =220$~ \kms), this expression shows that all globular clusters within a radius of 1~kpc have been lost due to dynamical friction, assuming an age of $t \simeq 12$~Gyr.

 The present-day number density profile of globular clusters in the Milky Way follows a power law with an index of about $-2.7$ in the range $R=1$--10~kpc. 
Because the profile within 1~kpc has been strongly affected by dynamical friction, to first order we can extrapolate this profile to $R=0$ to obtain an initial distribution. Under this assumption, a fraction of $(1/5)^{0.3}=0.62$ of all globular clusters within 5~kpc (where $\sim80\%$ of the metal-rich GCs reside) were formed within a radius of 1~kpc and were thus destroyed within a Hubble time.
For a mean cluster mass of $5~10^5$ \Msun\ the critical radius is $R_{\rm crit} = $0.64 kpc
and the fraction of clusters within 5 kpc that are destroyed is 0.54.
 Metal-poor globular clusters suffered a negligible decline due to their larger average galactocentric radii.\\
{\it We conclude that dynamical friction may have led to the destruction of about half of the clusters formed within $R_{\rm Gal} < 1$ kpc from the center thus explaining the small cluster-to-star ratio close to  the Galactic Center, but not the general trend that exists over a much larger distance range}.

\subsection{Cluster destruction by tidal shocks from the dense interstellar medium of high-redshift galaxies} \label{sec:shocks}

A considerable body of work has shown that stellar clusters are efficiently destroyed by (impulsive) tidal shocks from passing gas overdensities such as GMCs \citep[e.g.][]{lamers06a,gieles06,kruijssen11}, especially in the gas-rich environments at high redshift where globular clusters must have formed \citep{elmegreen10,kruijssen15b}. These works have shown that tidal shocks can be more than an order of magnitude more efficient at destroying clusters than tidal evaporation. The importance of tidal shocks from the ISM is underlined by the observation that there exists a strong correlation between the median cluster age and the ISM surface density and velocity dispersion in nearby galaxies \citep{miholics17}.

At face value, tidal shocks could explain the strong dependence of the cluster-to-star ratio on metallicity, because destruction processes provide a large dynamic range -- starting from a certain number of clusters per unit field-star mass, disruption can reduce that number by any amount until the Poisson limit of a single globular cluster is reached. In order to be a feasible explanation of the observed metallicity trends, the strength and/or duration of tidal shock-driven disruption must increase with the globular cluster metallicity. It has been proposed by \citet{kruijssen15b} that this is indeed the case, due to two effects. Firstly, the gas pressure increases towards higher galaxy masses (which have higher metallicities), resulting in increased cluster destruction towards high metallicities. The disruption time of a $10^5~\msun$ cluster is estimated to decrease from a few Gyr at ${\rm [Fe/H]}<-2$ to less than 100~Myr at ${\rm [Fe/H]}>-0.7$, with a total dynamic range of a factor of $\sim50$. Secondly, the duration of tidal-shock driven disruption is expected to weakly increase with metallicity, because the migration of clusters out of the gas-rich environment due to galaxy mergers occurs less frequently towards high galaxy masses. The duration is estimated to range from $0.1$--$1$~Gyr at ${\rm [Fe/H]}<-2$ to several Gyr at ${\rm [Fe/H]}>-0.7$. By taking the ratio between the disruption timescale and the migration timescale, we find that destruction is more efficient at high metallicities by a factor of 100--1000 than at low metallicities. 
This agrees very well with the observed trend of the cluster-to-star ratio decreasing by a factor of 100--1000 as a function of metallicity presented in \S\ref{sec:obs}.\footnote{At first sight, it may seem like the large (factor of 100--1000) range in disruption timescales as a function of metallicity should result in a strong relation between metallicity and the peak mass of the globular cluster mass function (GCMF), which is thought to have been shaped by cluster disruption \citep[e.g.][]{elmegreen97,fall01}. Such a relation is not observed \citep{jordan07}. However, it was shown by \citet[Figure~3]{kruijssen15b} that these large differences in disruption timescale lead to a range of GCMF peak masses at $z=0$ that only spans a factor of 3, in quantitative agreement with the observed dependence of the peak mass on galaxy mass \citep{jordan07}. This weakened metallicity dependence arises for three reasons. Firstly, cluster disruption by tidal shocks becomes highly inefficient at cluster masses $>10^6~\msun$, because the crossing time of the cluster becomes shorter than the typical duration of perturbations. As a result, the GCMF peak mass hardly increases any further due to disruption. Secondly, the initial cluster mass function rapidly steepens near the maximum cluster mass scale $M_{\rm c}$, which limits the maximum peak mass attainable by disruption. Thirdly, the remaining metallicity trend of the peak mass is largely erased during subsequent mass loss by evaporation in the galactic halo.}

\citet{kruijssen15b} further argued that the migration of globular clusters away from the gas-rich environment of their host galaxies proceeds through galaxy mergers, redistributing the cluster orbits in a way that largely erases their correlation to their birth environment, even if a weak metallicity gradient remains. This may explain why the cluster-to-star ratio depends strongly on the metallicity, but only weakly on the present-day galactocentric radius -- the current positions of globular clusters are only an indirect tracer of the high-redshift environment in which most of the disruption took place.

{\it The theoretical arguments outlined here suggest that the differences between the MDF of globular clusters and halo stars may be caused by tidal shock-driven disruption, which proceeds more rapidly and over a longer timescale in higher-metallicity galaxies. The implication of this is that the cluster-to-star ratio does not reflect a formation efficiency, but a survival fraction.}

\section{The survival of clusters in the cosmological context}
\label{sec:cosmology}

In many of the previous sections (all but \S\ref{sec:shocks}), we have assumed that the clusters are formed in their present host galaxies, and that their destruction depends on the present conditions in these host galaxies. However, these conditions may have changed during the cosmological history. 
Most generally, the hierarchical galaxy formation process means that present-day galaxies and their haloes have a large number of progenitors at the formation redshift of globular clusters \citep[e.g.][]{delucia07}. Therefore, the only way of understanding intra-galaxy variations of the cluster-to-star ratio at $z=0$ is by combining inter-galaxy variations at $z\sim3$, which indeed characterizes much of the discussion in this work. In this section we discuss the consequences of hierarchical galaxy growth for the survival of globular clusters.

The mass evolution of globular clusters during the cosmological growth of their host galaxies has traditionally been underemphasised, but recent studies are beginning to address the topic in more detail. \citet{prieto08} were the first to insert globular clusters into a dark matter-only simulation of hierarchical structure formation, following their evolution with a semi-analytic model. Their method was expanded by \citet{muratov10}, who added a simple chemical model to address the emergence of the colour bimodality of globular clusters.  \citet{tonini13} addressed the build up of globular cluster populations in massive galaxies through semi-analytic modeling using dark-matter only merger trees.  Although cluster formation and dissolution formation were not explicitly included in the simulations, the models showed that much of the metal poor globular cluster population in massive galaxies was mostly accreted from dwarf galaxies whereas the metal rich populations formed preferentially in-situ.  Hence, a considerable part of the globular clusters in the Milky Way and M31, possibly even most of the metal-poor clusters, may have originated in captured dwarf galaxies \citep{cote98,marin-franch09,lee07,mackey10,mackey13}. 
A similar scenario was proposed by \citet{hilker99} for the central Fornax cluster. 
So we have to consider their survival in dwarf galaxies for the first epoch of their lives and also where they are accreted into their new host galaxies.

\citet{kruijssen15b} has developed a semi-analytical model for the formation and evolution of globular cluster populations in a hierarchical galaxy formation context. In the following, we outline several of the main aspects of the model directly relevant to the MDFs of stars and clusters. In doing so, we distinguish between `accreted' and `in-situ' globular clusters, but we acknowledge that this represents a false dichotomy. In reality, galaxies form through hierarchical merging, implying that even the `main body' of a young galaxy is ill-defined and consists of several `ex-situ' progenitors. We use `in-situ' to refer to the globular clusters that formed during the initial collapse of the central galaxy, irrespective of the precise branch of its merger tree.

\subsection{Globular clusters formed in-situ} \label{sec:cosmology:insitu}

For the globular clusters that formed within the main-body of the young massive galaxy, the conditions that they experienced during their early evolution were very different than their current conditions (i.e. within the bulge/halo of the galaxy).  Given the mass-metallicity relation of galaxies, the (proto)massive galaxy is expected to be relatively metal rich, hence should preferentially form metal rich clusters.  As seen in local galaxies forming massive stellar clusters (YMCs), clusters form in the gas-rich discs of galaxies, even during major mergers \citep[e.g.][]{trancho07}, which likely also applies to globular clusters forming in the early Universe \citep[e.g.][]{kruijssen15b}.  The tidal perturbations from passing GMCs and the tidal field of the host galaxy can cause rapid mass loss amongst the young globular clusters, and if the globular clusters stay in the gas rich discs may completely dissolve.  Hence, the first few Gyr of a globular cluster's life may therefore be characterized by enhanced disruption \citep[e.g.][]{elmegreen10,kruijssen12c}.

However, if the young galaxy undergoes a merger, the young globular clusters can be liberated into the halo or bulge of the galaxy, away from the disruptive effects of GMCs \citep[e.g.][]{kruijssen12c}. This merger can be major or minor -- the globular clusters will escape the gas-rich environment as long as the host galaxy is merging with another galaxy of its own mass or higher. Given that the merger rate of galaxies was significantly higher than in the local Universe, this is expected to be an efficient mechanism in removing globular clusters from their birth-environment, i.e. outside the disk.  Once in the halo globular clusters evolve largely passively, losing mass due to two-body relaxation and (weaker) tidal perturbations (see \S\ref{sec:destruc}).

\subsection{Accreted globular clusters} \label{sec:cosmology:accreted}

As in the more massive spiral and early type galaxies, dwarf galaxies also host metallicity gradients \citep{tolstoy04, battaglia06, battaglia11, kirby11, kirby12}, which implies that a metal-rich population of globular clusters will be formed more concentrated towards the centre of a dwarf galaxy, i.e.~within a few kpc, than the old metal-poor population.  As discussed for the more massive galaxies, the metal rich clusters near the centre of the galaxy (i.e. those within $\lesssim2$~kpc) have a lower survival probability than those outside the central regions, meaning that cluster dissolution will push the star-to-cluster ratio to higher values for the metal rich populations.  In the outskirts of dwarfs, globular clusters are likely to survive for long periods meaning that their star-to-cluster ratios will be more reflective of their initial conditions than the central regions. However, the above works show that the range of metallicities of stars that formed in-situ within a single galaxy rarely exceeds more than an order of magnitude.\footnote{The metallicity gradient in the galactic disk, determined from Cepheids at $5 < R < 17$ kpc, is about -0.05 dex/kpc \citep{lemasle08} which implies a decrease in metallicity by only a factor 3 over a distance of 10 kpc.}  The more than 2 orders of magnitude in metallicity covered by globular cluster populations must therefore be obtained by merging galaxies of different masses.

When massive spirals/ellipticals accrete dwarf galaxies, they also accrete their globular cluster populations.  Where the globular clusters of dwarf galaxies end up within the more massive host depends on the location of the globular clusters within the dwarf at the time of accretion and also on the mass of the dwarf galaxy.  More massive dwarfs have larger binding energies, hence can survive the accretion onto the massive galaxy for a longer period.  This means that they are more likely to deposit their globular clusters closer to the centre of the massive galaxy.  Likewise, globular clusters on the outskirts of the dwarfs at the time of accretion (which are more likely to be metal poor) will be lost into the (outer) halo of the more massive galaxy.  Hence, the majority of metal-poor globular clusters from dwarf galaxies are likely (though not required) to be contributed to the (outer) halo of the more massive galaxy and, as a result, the dissolution parameter of the accreted globular clusters hardly changes during accretion \citep[e.g.][]{rieder13}. Indeed, as discussed in \S~\ref{sec:2.2.1} and \ref{sec:obssum}, the outer halos of the MW, M31 and Cen-A are similar to that of dwarf galaxies in terms of the MDFs of their stars and globular clusters.

\subsection{The hierarchical assembly of globular cluster systems and the role of metallicity} \label{sec:cosmology:hierarchy}

As discussed in \S\ref{sec:destruc}, we expect that most of the physics governing globular cluster formation and disruption were set by their environment at the time of formation. Specifically, the host galaxy mass at formation determined the formation efficiencies, maximum cluster masses, and (most importantly) the metallicity and cluster disruption timescales. We can no longer directly measure the mass of the galaxy in which a globular cluster formed. However, in the context of the above picture of hierarchical galaxy formation and globular cluster system assembly, we see that the globular cluster metallicities are good tracers of the masses of the host galaxies in which they formed \citep[e.g.][]{kruijssen15b}, thanks to the galaxy mass--metallicity relation, with a second-order dependence on the galactocentric radii in these galaxies. 

{\it We conclude that hierarchical galaxy growth weakens the relation between globular clusters and their (position within the) host galaxy, but the globular cluster metallicities are a good tracer of the host galaxy masses where globular clusters formed. Because the initial environment sets their formation and survival probability, it may not be surprising that the cluster-to-star ratio shows a clear trend with metallicity, that appears to be independent galactocentric distance in massive galaxies, and results in dissimilar MDFs of globular clusters and field stars.}

\section{Discussion  and conclusions}  
\label{sec:conclusions}

We have collected and reviewed information on the globular cluster-to-star ratios, \Rgcst, in different types of galaxies as a function of galactic mass and metallicity, and as a function of metallicity and location within galaxies. The data show the following:
\begin{enumerate}
\item the mean metallicity of galaxies is related to their mass, with the most massive galaxies having the highest metallicities; 
\item the cluster specific frequency, $S_N = N_{\rm cl}/M_{\rm stars}$, depends on galaxy mass with the highest values in the least massive galaxies; 
\item within a single galaxy, \Rgcst\ is a strong function of metallicity, with the highest ratio at low metallicity and the lowest ratio at high metallicity, and a total dynamic range as high as 2--3 orders of magnitude; 
\item one of the most extreme examples is the Fornax dSph, which has 5 clusters at $\feh\ <-1$ and no clusters at $\feh >-1$ although the stellar MDF peaks at $\feh=-1$. The dwarf galaxy IKN  with a stellar population of 
$\feh$ around -1.3 and one of it five clusters, viz. IKN-5 at -2.1 is another extreme example. 
\item due to a gradient of decreasing metallicity with galactocentric distance, \Rgcst\ is also a function of distance, with the smallest values near the centre, but it is unclear to what extent this dependence remains after accounting for the metallicity gradient; 
\end{enumerate}

We acknowledge that most of these trends were already found and discussed individually in the literature (see \S \ref{sec:obs}). They are repeated here because we want to show the overall picture presented by the observations and try to explain them in a coherent cosmological cluster evolution framework.

For the interpretation of these observations it is important to realize that under all star forming conditions, the fraction of the mass that ends up in massive clusters with $\Mi > 10^5~\Msun$ is small compared to the mass of the field stars, because most of the stars end up in unbound associations or in low mass clusters that are easily dissolved by tidal evaporation and tidal shocks. This implies that \Rgcst\  depends on the cluster formation efficiency, CFE, of massive clusters ($M > 10^5~\Msun$) and cluster destruction, and not on the star formation efficiency.

We discussed the various mechanisms that affect the cluster-to-star ratio. We first considered the dependence of the formation efficiency and early survival of clusters on the metallicity and environment.
\begin{enumerate}
\item[(i)] The cluster formation efficiency has a weak (and indirect) dependence on metallicity, but the predicted trend cannot explain the observations, because the CFE is expected to be higher in high metallicity environments, in contrast to the observed low cluster-to-star ratio in the inner halo (\S\ref{sec:form:cfe}).
\item[(ii)] `Infant mortality' of young clusters by gas expulsion cannot explain the observations, because in the dense environment where massive clusters are born, the amount of gas in young globular clusters is too small to unbind the clusters (\S\ref{sec:form:expulsion}).
\item[(iii)] Because GCs are typically massive, $M > 10^4 ~\Msun $, the formation efficiency of GCs depends on the maximum mass of clusters that can be formed in various environments. Theoretical arguments, supported by observational evidence, suggest that the maximum cluster mass increases with increasing mass of giant molecular clouds, which depends on the mass and density of the galaxy and hence on the metallicity. Therefore the formation efficiency of GCs is expected to increase with increasing metallicity. This is contrary to the observed decrease of \Rgcst\ with increasing \feh\ (\S\ref{sec:form:mmax}).
\end{enumerate}

Considering that cluster formation does not seem to explain the observed trends, we also discussed the destruction of GCs in various environments.
\begin{enumerate}
\item[(iv)]  Dissolution of clusters by tidal evaporation depends on the galactic potential at the distance of their orbits. In principle, clusters in small orbits are destroyed faster than clusters in wide orbits. If the GCs are formed in-situ in galaxies with a radial metallicity gradient, the metal rich clusters will be destroyed more easily than metal poor clusters on wider orbits. However, we found that tidal evaporation alone can only destroy GCs within a Hubble time if they are on
circular orbits within $\sim1$ kpc (unless they are on highly elliptical orbits). This may explain the lack of metal rich GCs
in the very centre of massive galaxies. However the observed trend of \Rgcst\ with metallicity extends over a much wider  distance range, implying that tidal evaporation cannot be responsible (\S \ref{sec:stripping}).
\item[(v)]  Dynamical friction leads to the inspiraling of clusters towards the galactic centre. Contrary to dissolution, the lifetimes of clusters 
for spiraling into the galactic centre decreases with mass, $t_{\rm df} \propto \Rgal^2\times M^{-1}$. When clusters spiral in to very small orbits, they will be destroyed by tidal evaporation and shocks. We quantified the effect due to evaporation in typical spiral galaxies and found that it will reduce the surviving number of massive clusters with $M \simeq\ 5~10^5~\Msun$ on orbits smaller than about 1 kpc. This may explain the small cluster-to-star ratios of metal-rich clusters observed in the center of massive galaxies. However, the observed trend with metallicity extends over a much wider distance range (\S \ref{sec:dynfric}).
\item[(vi)] Clusters can also be destroyed by tidal shocks. In gas-rich galaxies at high redshift, where GCs must have formed, the destruction of GCs by tidal shocks can be orders of magnitude more efficient than tidal evaporation. We studied the environmental conditions during GC formation and found that shock-driven destruction in the environment where metal-rich clusters are formed is 100-1000 times more efficient than in the environment where metal-poor GCs are formed.
This quantitatively agrees with the observed large range of \Rgcst\ with metallicity. It also shows that \Rgcst\ does not reflect the cluster formation efficiency, but the survival probability (\S \ref{sec:shocks}).
\end{enumerate}

Since galaxies are formed hierarchically and the present day surviving GCs have followed the history of the galaxies
in which they we formed originally, we also considered the fate of clusters in the context of hierarchical galaxy formation.
\begin{enumerate}
\item[(vii)] GCs that formed in-situ in the high density environment of gas-rich disks may leave their shock-destructive 
formation environment during a merger with a galaxy of the same or higher mass. This is an efficient mechanism to move GCs out of the disk. Once they are in the halo, the GCs evolve passively and lose mass inefficiently by tidal evaporation
(\S \ref{sec:cosmology:insitu}).
\item[(viii)] The radial metallicity gradients in present-day spirals covers only a factor $\sim 10$, which is much smaller than the
range in metallicity of the GCs in a galaxy, $\sim100$ to 1000. Therefore the metal-poor GCs must have  
been captured from low-metallicity dwarf galaxies during the hierarchical growth of the central galaxy. Low-mass dwarf galaxies
with  low-metallicity GCs have a small binding energy, causing their clusters to be  released in the low-density 
halo of the capturing galaxy.  On the other hand, more massive dwarfs with more metal-rich GCs have stronger binding energy and will release their GCs deeper into the capturing galaxy. This explains, at least qualitatively, the radial metallicity gradient of GCs as a reflection of the host galaxy mass at GC formation (\S \ref{sec:cosmology:accreted}).
\item[(ix)] The merging of galaxies and the capture of GCs implies that the present-day 
location of the GCs is not a good indicator of their formation environment. However, combining the results presented 
above, in particular conclusions (vi), (vii) and (viii), imply that the \Rgcst\ is expected to depend strongly on metallicity
and only weakly on galactocentric distance in their present host galaxy. This is qualitatively in agreement with the 
observed trends (\S \ref{sec:cosmology:hierarchy}).
\end{enumerate}

When we discussed the observations in \S\ref{sec:obssum} we posed five specific questions. We can now tentatively 
try to answer these.
\begin{enumerate}
\item[Q1.] GCs in dwarf galaxies are more metal-poor than the stars, because the metal-rich clusters are formed in high density 
environments where destruction by shocks is efficient. 
This may be exacerbated by continued star and (low-mass) cluster formation in dwarfs that added metal-rich stars but no metal-rich globular clusters.\footnote{This effect is most important in dwarf galaxies, because there we cannot separate the halo field stars from the younger disc field stars.}
\item[Q2.] The same arguments apply to the inner regions of the halos of large galaxies. Metal-rich clusters formed in-situ
will be destroyed by shocks in their original environment. In addition, dynamical friction may have led to the destruction of the innermost clusters.
\item[Q3.] The metal-poor clusters in the outer halos of large galaxies are captured from low-mass dwarfs that had a low binding energy. The destruction rate in these dwarf galaxies and in the outer halo is small, so these clusters have a higher survival probability.
\item[Q4.] The strong dependence of the cluster-to-star ratio on metallicity is the result of the fact that the low-metallicity clusters were formed in dwarfs where the shock destruction is not efficient, whereas the more metal-rich clusters are formed in more massive galaxies (because metallicity increases with galaxy mass) where the environment is more destructive.
\item[Q5.] The present day location of GCs is only weakly related to their original location and environment. The dominant factor that sets the survival probability of clusters is the metallicity (via the relation between metallicity and galaxy mass). Hierarchical galaxy growth weakens initial radial trends. That is why 
the relation between \Rgcst\ and \feh\ is almost independent of distance.
\end{enumerate}

Many of the effects proposed in this paper are only described qualitatively. In future work, we intend to quantify the framework sketched in this work using self-consistent numerical simulations of globular cluster formation and evolution during galaxy formation (Pfeffer et al.~in prep., Kruijssen et al.~in prep.). Such simulations will allow us to directly follow the evolution of the cluster and field star MDFs while galaxies are forming and the cluster population is evolving due to the ongoing formation and destruction of stellar clusters.

\begin{acknowledgements}

The authors thank the referee, S\o ren Larsen, and Bill Harris for extensive and constructive feedback that greatly improved this work.
 HJGLML thanks ESO for a Visiting Scientist Fellowship at Garching, the LKBF for a travel grant to Liverpool, and Heidelberg University for a visiting scientist grant through Sonderforschungsbereich SFB 881 ``The Milky Way System'' (subproject P1) of the German Research Foundation (DFG). JMDK gratefully acknowledges funding from the German Research Foundation (DFG) in the form of an Emmy Noether Research Group (grant number KR4801/1-1, PI Kruijssen), and from the European Research Council (ERC) under the European Union's Horizon 2020 research and innovation programme via the ERC Starting Grant MUSTANG (grant agreement number 714907, PI Kruijssen). NB is partially funded by a Royal Society University Research Fellowship and a European Research Council (ERC) Consolidator Grant (Multi-Pop, grant agreement number 646928).

\end{acknowledgements}

\bibliographystyle{aa}

\bibliography{mybib}

\begin{thebibliography}{158}
\expandafter\ifx\csname natexlab\endcsname\relax\def\natexlab#1{#1}\fi

\bibitem[{{Adamo} {et~al.}(2015){Adamo}, {Kruijssen}, {Bastian}, {Silva-Villa},
  \& {Ryon}}]{adamo15b}
{Adamo}, A., {Kruijssen}, J.~M.~D., {Bastian}, N., {Silva-Villa}, E., \&
  {Ryon}, J. 2015, \mnras, 452, 246

\bibitem[{{An} {et~al.}(2012){An}, {Beers}, {Johnson}, {Pinsonneault}, {Lee},
  {Ivezi{\'c}}, \& {Newby}}]{an12}
{An}, D., {Beers}, T.~C., {Johnson}, J.~A., {et~al.} 2012, ASPC, 458, 179

\bibitem[{{Ashman} \& {Zepf}(1992)}]{ashman92}
{Ashman}, K.~M. \& {Zepf}, S.~E. 1992, \apj, 384, 50

\bibitem[{{Barker} {et~al.}(2012){Barker}, {Ferguson}, {Irwin}, {Arimoto}, \&
  {Jablonka}}]{barker12}
{Barker}, M.~K., {Ferguson}, A.~M.~N., {Irwin}, M.~J., {Arimoto}, N., \&
  {Jablonka}, P. 2012, \mnras, 419, 1489

\bibitem[{{Bastian}(2008)}]{bastian08}
{Bastian}, N. 2008, \mnras, 390, 759

\bibitem[{{Battaglia} {et~al.}(2011){Battaglia}, {Tolstoy}, {Helmi}, {Irwin},
  {Parisi}, {Hill}, \& {Jablonka}}]{battaglia11}
{Battaglia}, G., {Tolstoy}, E., {Helmi}, A., {et~al.} 2011, \mnras, 411, 1013

\bibitem[{{Battaglia} {et~al.}(2006){Battaglia}, {Tolstoy}, {Helmi}, {Irwin},
  {Letarte}, {Jablonka}, {Hill}, {Venn}, {Shetrone}, {Arimoto}, {Primas},
  {Kaufer}, {Francois}, {Szeifert}, {Abel}, \& {Sadakane}}]{battaglia06}
{Battaglia}, G., {Tolstoy}, E., {Helmi}, A., {et~al.} 2006, \aap, 459, 423

\bibitem[{{Baumgardt} \& {Kroupa}(2007)}]{baumgardt07}
{Baumgardt}, H. \& {Kroupa}, P. 2007, \mnras, 380, 1589

\bibitem[{{Baumgardt} \& {Makino}(2003)}]{baumgardt03}
{Baumgardt}, H. \& {Makino}, J. 2003, \mnras, 340, 227

\bibitem[{{Beasley} {et~al.}(2008){Beasley}, {Bridges}, {Peng}, {Harris},
  {Harris}, {Forbes}, \& {Mackie}}]{beasley08}
{Beasley}, M.~A., {Bridges}, T., {Peng}, E., {et~al.} 2008, \mnras, 386, 1443

\bibitem[{{Bekki} {et~al.}(2002){Bekki}, {Forbes}, {Beasley}, \&
  {Couch}}]{bekki02}
{Bekki}, K., {Forbes}, D.~A., {Beasley}, M.~A., \& {Couch}, W.~J. 2002, \mnras,
  335, 1176

\bibitem[{{Binney} \& {Tremaine}(1987)}]{binney87}
{Binney}, J. \& {Tremaine}, S. 1987, {Galactic dynamics} (Princeton, NJ,
  Princeton University Press, 1987, 747 pp.)

\bibitem[{{Boily} \& {Kroupa}(2003)}]{boily03}
{Boily}, C.~M. \& {Kroupa}, P. 2003, \mnras, 338, 665

\bibitem[{{Brodie} \& {Strader}(2006)}]{brodie06}
{Brodie}, J.~P. \& {Strader}, J. 2006, \araa, 44, 193

\bibitem[{{Brodie} {et~al.}(2012){Brodie}, {Usher}, {Conroy}, {Strader},
  {Arnold}, {Forbes}, \& {Romanowsky}}]{brodie12}
{Brodie}, J.~P., {Usher}, C., {Conroy}, C., {et~al.} 2012, \apjl, 759, L33

\bibitem[{{Burgarella} {et~al.}(2001){Burgarella}, {Kissler-Patig}, \&
  {Buat}}]{burgarella01}
{Burgarella}, D., {Kissler-Patig}, M., \& {Buat}, V. 2001, \aj, 121, 2647

\bibitem[{{Capriotti} \& {Hawley}(1996)}]{capriotti96}
{Capriotti}, E.~R. \& {Hawley}, S.~L. 1996, \apj, 464, 765

\bibitem[{{Carollo} {et~al.}(2010){Carollo}, {Beers}, {Chiba}, {Norris},
  {Freeman}, {Lee}, {Ivezi{\'c}}, {Rockosi}, \& {Yanny}}]{carollo10}
{Carollo}, D., {Beers}, T.~C., {Chiba}, M., {et~al.} 2010, \apj, 712, 692

\bibitem[{{Chandrasekhar}(1943)}]{chandrasekhar43}
{Chandrasekhar}, S. 1943, \apj, 97, 255

\bibitem[{{Chapman} {et~al.}(2006){Chapman}, {Ibata}, {Lewis}, {Ferguson},
  {Irwin}, {McConnachie}, \& {Tanvir}}]{chapman06}
{Chapman}, S.~C., {Ibata}, R., {Lewis}, G.~F., {et~al.} 2006, \apj, 653, 255

\bibitem[{{Cole} {et~al.}(2000){Cole}, {Lacey}, {Baugh}, \& {Frenk}}]{cole00}
{Cole}, S., {Lacey}, C.~G., {Baugh}, C.~M., \& {Frenk}, C.~S. 2000, \mnras,
  319, 168

\bibitem[{{Cote} {et~al.}(1998){Cote}, {Marzke}, \& {West}}]{cote98}
{Cote}, P., {Marzke}, R.~O., \& {West}, M.~J. 1998, \apj, 501, 554

\bibitem[{{Cottaar} {et~al.}(2012){Cottaar}, {Meyer}, {Andersen}, \&
  {Espinoza}}]{cottaar12}
{Cottaar}, M., {Meyer}, M.~R., {Andersen}, M., \& {Espinoza}, P. 2012, \aap,
  539, A5

\bibitem[{{Crnojevi{\'c}} {et~al.}(2013){Crnojevi{\'c}}, {Ferguson}, {Irwin},
  {Bernard}, {Arimoto}, {Jablonka}, \& {Kobayashi}}]{crnojevic13}
{Crnojevi{\'c}}, D., {Ferguson}, A.~M.~N., {Irwin}, M.~J., {et~al.} 2013,
  \mnras, 432, 832

\bibitem[{{Crnojevi{\'c}} {et~al.}(2010){Crnojevi{\'c}}, {Grebel}, \&
  {Koch}}]{crnojevic10}
{Crnojevi{\'c}}, D., {Grebel}, E.~K., \& {Koch}, A. 2010, \aap, 516, A85

\bibitem[{{De Lucia} \& {Blaizot}(2007)}]{delucia07}
{De Lucia}, G. \& {Blaizot}, J. 2007, \mnras, 375, 2

\bibitem[{{Dessauges-Zavadsky} {et~al.}(2017){Dessauges-Zavadsky}, {Schaerer},
  {Cava}, {Mayer}, \& {Tamburello}}]{dessauges17}
{Dessauges-Zavadsky}, M., {Schaerer}, D., {Cava}, A., {Mayer}, L., \&
  {Tamburello}, V. 2017, \apjl, 836, L22

\bibitem[{{Durrell} {et~al.}(2001){Durrell}, {Harris}, \&
  {Pritchet}}]{durrell01}
{Durrell}, P.~R., {Harris}, W.~E., \& {Pritchet}, C.~J. 2001, \aj, 121, 2557

\bibitem[{{Durrell} {et~al.}(2004){Durrell}, {Harris}, \&
  {Pritchet}}]{durrell04}
{Durrell}, P.~R., {Harris}, W.~E., \& {Pritchet}, C.~J. 2004, \aj, 128, 260

\bibitem[{{Durrell} {et~al.}(2010){Durrell}, {Sarajedini}, \&
  {Chandar}}]{durrell10}
{Durrell}, P.~R., {Sarajedini}, A., \& {Chandar}, R. 2010, \apj, 718, 1118

\bibitem[{{Elmegreen}(2010)}]{elmegreen10}
{Elmegreen}, B.~G. 2010, \apjl, 712, L184

\bibitem[{{Elmegreen} \& {Efremov}(1997)}]{elmegreen97}
{Elmegreen}, B.~G. \& {Efremov}, Y.~N. 1997, \apj, 480, 235

\bibitem[{{Erb} {et~al.}(2006){Erb}, {Steidel}, {Shapley}, {Pettini}, {Reddy},
  \& {Adelberger}}]{erb06}
{Erb}, D.~K., {Steidel}, C.~C., {Shapley}, A.~E., {et~al.} 2006, \apj, 646, 107

\bibitem[{{Fall} \& {Zhang}(2001)}]{fall01}
{Fall}, S.~M. \& {Zhang}, Q. 2001, \apj, 561, 751

\bibitem[{{Forbes} \& {Bridges}(2010)}]{forbes10}
{Forbes}, D.~A. \& {Bridges}, T. 2010, \mnras, 404, 1203

\bibitem[{{Forbes} {et~al.}(1997){Forbes}, {Brodie}, \& {Grillmair}}]{forbes97}
{Forbes}, D.~A., {Brodie}, J.~P., \& {Grillmair}, C.~J. 1997, \aj, 113, 1652

\bibitem[{{F{\"o}rster Schreiber} {et~al.}(2009){F{\"o}rster Schreiber},
  {Genzel}, {Bouch{\'e}}, {Cresci}, {Davies}, {Buschkamp}, {Shapiro},
  {Tacconi}, {Hicks}, {Genel}, {Shapley}, {Erb}, {Steidel}, {Lutz},
  {Eisenhauer}, {Gillessen}, {Sternberg}, {Renzini}, {Cimatti}, {Daddi},
  {Kurk}, {Lilly}, {Kong}, {Lehnert}, {Nesvadba}, {Verma}, {McCracken},
  {Arimoto}, {Mignoli}, \& {Onodera}}]{forsterschreiber09}
{F{\"o}rster Schreiber}, N.~M., {Genzel}, R., {Bouch{\'e}}, N., {et~al.} 2009,
  \apj, 706, 1364

\bibitem[{{Frebel} \& {Norris}(2015)}]{frebel15}
{Frebel}, A. \& {Norris}, J.~E. 2015, \araa, 53, 631

\bibitem[{{Freeman} {et~al.}(2017){Freeman}, {Rosolowsky}, {Kruijssen},
  {Bastian}, \& {Adamo}}]{freeman17}
{Freeman}, P., {Rosolowsky}, E., {Kruijssen}, J.~M.~D., {Bastian}, N., \&
  {Adamo}, A. 2017, \mnras, 468, 1769

\bibitem[{{Gallazzi} {et~al.}(2005){Gallazzi}, {Charlot}, {Brinchmann},
  {White}, \& {Tremonti}}]{gallazzi05}
{Gallazzi}, A., {Charlot}, S., {Brinchmann}, J., {White}, S.~D.~M., \&
  {Tremonti}, C.~A. 2005, \mnras, 362, 41

\bibitem[{{Galleti} {et~al.}(2009){Galleti}, {Bellazzini}, {Buzzoni},
  {Federici}, \& {Fusi Pecci}}]{galleti09}
{Galleti}, S., {Bellazzini}, M., {Buzzoni}, A., {Federici}, L., \& {Fusi
  Pecci}, F. 2009, \aap, 508, 1285

\bibitem[{{Galleti} {et~al.}(2004){Galleti}, {Federici}, {Bellazzini}, {Fusi
  Pecci}, \& {Macrina}}]{galleti04}
{Galleti}, S., {Federici}, L., {Bellazzini}, M., {Fusi Pecci}, F., \&
  {Macrina}, S. 2004, \aap, 416, 917

\bibitem[{{Gebhardt} \& {Kissler-Patig}(1999)}]{gebhardt99}
{Gebhardt}, K. \& {Kissler-Patig}, M. 1999, \aj, 118, 1526

\bibitem[{{Genzel} {et~al.}(2011){Genzel}, {Newman}, {Jones}, {F{\"o}rster
  Schreiber}, {Shapiro}, {Genel}, {Lilly}, {Renzini}, {Tacconi}, {Bouch{\'e}},
  {Burkert}, {Cresci}, {Buschkamp}, {Carollo}, {Ceverino}, {Davies}, {Dekel},
  {Eisenhauer}, {Hicks}, {Kurk}, {Lutz}, {Mancini}, {Naab}, {Peng},
  {Sternberg}, {Vergani}, \& {Zamorani}}]{genzel11}
{Genzel}, R., {Newman}, S., {Jones}, T., {et~al.} 2011, \apj, 733, 101

\bibitem[{{Georgiev} {et~al.}(2010){Georgiev}, {Puzia}, {Goudfrooij}, \&
  {Hilker}}]{georgiev10}
{Georgiev}, I.~Y., {Puzia}, T.~H., {Goudfrooij}, P., \& {Hilker}, M. 2010,
  \mnras, 406, 1967

\bibitem[{{Georgiev} {et~al.}(2009){Georgiev}, {Puzia}, {Hilker}, \&
  {Goudfrooij}}]{georgiev09}
{Georgiev}, I.~Y., {Puzia}, T.~H., {Hilker}, M., \& {Goudfrooij}, P. 2009,
  \mnras, 392, 879

\bibitem[{{Gieles} {et~al.}(2006{\natexlab{a}}){Gieles}, {Larsen}, {Anders},
  {Bastian}, \& {Stein}}]{gieles06a}
{Gieles}, M., {Larsen}, S.~S., {Anders}, P., {Bastian}, N., \& {Stein}, I.~T.
  2006{\natexlab{a}}, \aap, 450, 129

\bibitem[{{Gieles} {et~al.}(2006{\natexlab{b}}){Gieles}, {Portegies Zwart},
  {Baumgardt}, {Athanassoula}, {Lamers}, {Sipior}, \& {Leenaarts}}]{gieles06}
{Gieles}, M., {Portegies Zwart}, S.~F., {Baumgardt}, H., {et~al.}
  2006{\natexlab{b}}, \mnras, 371, 793

\bibitem[{{Gilbert} {et~al.}(2014){Gilbert}, {Kalirai}, {Guhathakurta},
  {Beaton}, {Geha}, {Kirby}, {Majewski}, {Patterson}, {Tollerud}, {Bullock},
  {Tanaka}, \& {Chiba}}]{gilbert14}
{Gilbert}, K.~M., {Kalirai}, J.~S., {Guhathakurta}, P., {et~al.} 2014, \apj,
  796, 76

\bibitem[{{Girichidis} {et~al.}(2012){Girichidis}, {Federrath}, {Allison},
  {Banerjee}, \& {Klessen}}]{girichidis12b}
{Girichidis}, P., {Federrath}, C., {Allison}, R., {Banerjee}, R., \& {Klessen},
  R.~S. 2012, \mnras, 420, 3264

\bibitem[{{Goddard} {et~al.}(2010){Goddard}, {Bastian}, \&
  {Kennicutt}}]{goddard10}
{Goddard}, Q.~E., {Bastian}, N., \& {Kennicutt}, R.~C. 2010, \mnras, 405, 857

\bibitem[{{Gonzalez} {et~al.}(2011){Gonzalez}, {Rejkuba}, {Zoccali}, {Valenti},
  \& {Minniti}}]{gonzalez11}
{Gonzalez}, O.~A., {Rejkuba}, M., {Zoccali}, M., {Valenti}, E., \& {Minniti},
  D. 2011, \aap, 534, A3

\bibitem[{{Gratton} {et~al.}(2012){Gratton}, {Carretta}, \&
  {Bragaglia}}]{gratton12}
{Gratton}, R.~G., {Carretta}, E., \& {Bragaglia}, A. 2012, \aapr, 20, 50

\bibitem[{{Grebel} {et~al.}(2003){Grebel}, {Gallagher}, \&
  {Harbeck}}]{grebel03}
{Grebel}, E.~K., {Gallagher}, III, J.~S., \& {Harbeck}, D. 2003, \aj, 125, 1926

\bibitem[{{Gregersen} {et~al.}(2015){Gregersen}, {Seth}, {Williams}, {Lang},
  {Dalcanton}, {Girardi}, {Skillman}, {Bell}, {Dolphin}, {Fouesneau},
  {Guhathakurta}, {Hamren}, {Johnson}, {Kalirai}, {Lewis}, {Monachesi}, \&
  {Olsen}}]{gregersen15}
{Gregersen}, D., {Seth}, A.~C., {Williams}, B.~F., {et~al.} 2015, \aj, 150, 189

\bibitem[{{Grillmair} {et~al.}(1996){Grillmair}, {Lauer}, {Worthey}, {Faber},
  {Freedman}, {Madore}, {Ajhar}, {Baum}, {Holtzman}, {Lynds}, {O'Neil}, \&
  {Stetson}}]{grillmair96}
{Grillmair}, C.~J., {Lauer}, T.~R., {Worthey}, G., {et~al.} 1996, \aj, 112,
  1975

\bibitem[{{Harmsen} {et~al.}(2017){Harmsen}, {Monachesi}, {Bell}, {de Jong},
  {Bailin}, {Radburn-Smith}, \& {Holwerda}}]{harmsen17}
{Harmsen}, B., {Monachesi}, A., {Bell}, E.~F., {et~al.} 2017, \mnras, 466, 1491

\bibitem[{{Harris} \& {Harris}(2000)}]{harris00}
{Harris}, G.~L.~H. \& {Harris}, W.~E. 2000, \aj, 120, 2423

\bibitem[{{Harris} {et~al.}(1999){Harris}, {Harris}, \& {Poole}}]{harris99}
{Harris}, G.~L.~H., {Harris}, W.~E., \& {Poole}, G.~B. 1999, \aj, 117, 855

\bibitem[{{Harris} {et~al.}(2010){Harris}, {Rejkuba}, \& {Harris}}]{harris10}
{Harris}, G.~L.~H., {Rejkuba}, M., \& {Harris}, W.~E. 2010, \pasa, 27, 457

\bibitem[{{Harris}(1991)}]{harris91}
{Harris}, W.~E. 1991, \araa, 29, 543

\bibitem[{{Harris}(1996)}]{harris96}
{Harris}, W.~E. 1996, \aj, 112, 1487

\bibitem[{{Harris}(2001)}]{harris01a}
{Harris}, W.~E. 2001, in Saas-Fee Advanced Course 28: Star Clusters, ed.
  L.~{Labhardt} \& B.~{Binggeli}, 223--+

\bibitem[{{Harris}(2003)}]{harris03}
{Harris}, W.~E. 2003, in Extragalactic Globular Cluster Systems, ed.
  M.~{Kissler-Patig}, 317

\bibitem[{{Harris} {et~al.}(2016){Harris}, {Blakeslee}, {Whitmore}, {Gnedin},
  {Geisler}, \& {Rothberg}}]{harris16}
{Harris}, W.~E., {Blakeslee}, J.~P., {Whitmore}, B.~C., {et~al.} 2016, \apj,
  817, 58

\bibitem[{{Harris} {et~al.}(2017){Harris}, {Ciccone}, {Eadie}, {Gnedin},
  {Geisler}, {Rothberg}, \& {Bailin}}]{harris17a}
{Harris}, W.~E., {Ciccone}, S.~M., {Eadie}, G.~M., {et~al.} 2017, \apj, 835,
  101

\bibitem[{{Harris} \& {Harris}(2001)}]{harrisharris01}
{Harris}, W.~E. \& {Harris}, G.~L.~H. 2001, \aj, 122, 3065

\bibitem[{{Harris} \& {Harris}(2002)}]{harrisharris02}
{Harris}, W.~E. \& {Harris}, G.~L.~H. 2002, \aj, 123, 3108

\bibitem[{{Harris} {et~al.}(2007{\natexlab{a}}){Harris}, {Harris}, {Layden}, \&
  {Stetson}}]{harris07b}
{Harris}, W.~E., {Harris}, G.~L.~H., {Layden}, A.~C., \& {Stetson}, P.~B.
  2007{\natexlab{a}}, \aj, 134, 43

\bibitem[{{Harris} {et~al.}(2007{\natexlab{b}}){Harris}, {Harris}, {Layden}, \&
  {Wehner}}]{harris07a}
{Harris}, W.~E., {Harris}, G.~L.~H., {Layden}, A.~C., \& {Wehner}, E.~M.~H.
  2007{\natexlab{b}}, \apj, 666, 903

\bibitem[{{Harris} \& {van den Bergh}(1981)}]{harris81}
{Harris}, W.~E. \& {van den Bergh}, S. 1981, \aj, 86, 1627

\bibitem[{{Haschke} {et~al.}(2012){Haschke}, {Grebel}, {Duffau}, \&
  {Jin}}]{haschke12}
{Haschke}, R., {Grebel}, E.~K., {Duffau}, S., \& {Jin}, S. 2012, \aj, 143, 48

\bibitem[{{Helmi} {et~al.}(2006){Helmi}, {Irwin}, {Tolstoy}, {Battaglia},
  {Hill}, {Jablonka}, {Venn}, {Shetrone}, {Letarte}, {Arimoto}, {Abel},
  {Francois}, {Kaufer}, {Primas}, {Sadakane}, \& {Szeifert}}]{helmi06}
{Helmi}, A., {Irwin}, M.~J., {Tolstoy}, E., {et~al.} 2006, \apjl, 651, L121

\bibitem[{{Hilker} {et~al.}(1999){Hilker}, {Infante}, \& {Richtler}}]{hilker99}
{Hilker}, M., {Infante}, L., \& {Richtler}, T. 1999, \aaps, 138, 55

\bibitem[{{Ibata} {et~al.}(2014){Ibata}, {Lewis}, {McConnachie}, {Martin},
  {Irwin}, {Ferguson}, {Babul}, {Bernard}, {Chapman}, {Collins}, {Fardal},
  {Mackey}, {Navarro}, {Pe{\~n}arrubia}, {Rich}, {Tanvir}, \&
  {Widrow}}]{ibata14}
{Ibata}, R.~A., {Lewis}, G.~F., {McConnachie}, A.~W., {et~al.} 2014, \apj, 780,
  128

\bibitem[{{Johnson} {et~al.}(2016){Johnson}, {Seth}, {Dalcanton}, {Beerman},
  {Fouesneau}, {Lewis}, {Weisz}, {Williams}, {Bell}, {Dolphin}, {Larsen},
  {Sandstrom}, \& {Skillman}}]{johnson16}
{Johnson}, L.~C., {Seth}, A.~C., {Dalcanton}, J.~J., {et~al.} 2016, \apj, 827,
  33

\bibitem[{{Johnson} {et~al.}(2017){Johnson}, {Seth}, {Dalcanton}, {Beerman},
  {Fouesneau}, {Weisz}, {Bell}, {Dolphin}, {Sandstrom}, \&
  {Williams}}]{johnson17}
{Johnson}, L.~C., {Seth}, A.~C., {Dalcanton}, J.~J., {et~al.} 2017, \apj, 839,
  78

\bibitem[{{Jord{\'a}n} {et~al.}(2007){Jord{\'a}n}, {McLaughlin},
  {C{\^o}t{\'e}}, {Ferrarese}, {Peng}, {Mei}, {Villegas}, {Merritt}, {Tonry},
  \& {West}}]{jordan07}
{Jord{\'a}n}, A., {McLaughlin}, D.~E., {C{\^o}t{\'e}}, P., {et~al.} 2007,
  \apjs, 171, 101

\bibitem[{{Juri{\'c}} {et~al.}(2008){Juri{\'c}}, {Ivezi{\'c}}, {Brooks},
  {Lupton}, {Schlegel}, {Finkbeiner}, {Padmanabhan}, {Bond}, {Sesar},
  {Rockosi}, {Knapp}, {Gunn}, {Sumi}, {Schneider}, {Barentine}, {Brewington},
  {Brinkmann}, {Fukugita}, {Harvanek}, {Kleinman}, {Krzesinski}, {Long},
  {Neilsen}, {Nitta}, {Snedden}, \& {York}}]{juric08}
{Juri{\'c}}, M., {Ivezi{\'c}}, {\v Z}., {Brooks}, A., {et~al.} 2008, \apj, 673,
  864

\bibitem[{{Kalirai} {et~al.}(2006){Kalirai}, {Gilbert}, {Guhathakurta},
  {Majewski}, {Ostheimer}, {Rich}, {Cooper}, {Reitzel}, \&
  {Patterson}}]{kalirai06}
{Kalirai}, J.~S., {Gilbert}, K.~M., {Guhathakurta}, P., {et~al.} 2006, \apj,
  648, 389

\bibitem[{{Kirby} {et~al.}(2012){Kirby}, {Cohen}, \& {Bellazzini}}]{kirby12}
{Kirby}, E.~N., {Cohen}, J.~G., \& {Bellazzini}, M. 2012, \apj, 751, 46

\bibitem[{{Kirby} {et~al.}(2013){Kirby}, {Cohen}, {Guhathakurta}, {Cheng},
  {Bullock}, \& {Gallazzi}}]{kirby13}
{Kirby}, E.~N., {Cohen}, J.~G., {Guhathakurta}, P., {et~al.} 2013, \apj, 779,
  102

\bibitem[{{Kirby} {et~al.}(2011){Kirby}, {Lanfranchi}, {Simon}, {Cohen}, \&
  {Guhathakurta}}]{kirby11}
{Kirby}, E.~N., {Lanfranchi}, G.~A., {Simon}, J.~D., {Cohen}, J.~G., \&
  {Guhathakurta}, P. 2011, \apj, 727, 78

\bibitem[{{Kruijssen}(2012)}]{kruijssen12d}
{Kruijssen}, J.~M.~D. 2012, \mnras, 426, 3008

\bibitem[{{Kruijssen}(2014)}]{kruijssen14c}
{Kruijssen}, J.~M.~D. 2014, Classical and Quantum Gravity, 31, 244006

\bibitem[{{Kruijssen}(2015)}]{kruijssen15b}
{Kruijssen}, J.~M.~D. 2015, \mnras, 454, 1658

\bibitem[{{Kruijssen} {et~al.}(2012{\natexlab{a}}){Kruijssen}, {Maschberger},
  {Moeckel}, {Clarke}, {Bastian}, \& {Bonnell}}]{kruijssen12}
{Kruijssen}, J.~M.~D., {Maschberger}, T., {Moeckel}, N., {et~al.}
  2012{\natexlab{a}}, \mnras, 419, 841

\bibitem[{{Kruijssen} {et~al.}(2012{\natexlab{b}}){Kruijssen}, {Pelupessy},
  {Lamers}, {Portegies Zwart}, {Bastian}, \& {Icke}}]{kruijssen12c}
{Kruijssen}, J.~M.~D., {Pelupessy}, F.~I., {Lamers}, H.~J.~G.~L.~M., {et~al.}
  2012{\natexlab{b}}, \mnras, 421, 1927

\bibitem[{{Kruijssen} {et~al.}(2011){Kruijssen}, {Pelupessy}, {Lamers},
  {Portegies Zwart}, \& {Icke}}]{kruijssen11}
{Kruijssen}, J.~M.~D., {Pelupessy}, F.~I., {Lamers}, H.~J.~G.~L.~M., {Portegies
  Zwart}, S.~F., \& {Icke}, V. 2011, \mnras, 414, 1339

\bibitem[{{Kruijssen} \& {Portegies Zwart}(2009)}]{kruijssen09b}
{Kruijssen}, J.~M.~D. \& {Portegies Zwart}, S.~F. 2009, \apjl, 698, L158

\bibitem[{{Lada} {et~al.}(1984){Lada}, {Margulis}, \& {Dearborn}}]{lada84}
{Lada}, C.~J., {Margulis}, M., \& {Dearborn}, D. 1984, \apj, 285, 141

\bibitem[{{Lamers} {et~al.}(2010){Lamers}, {Baumgardt}, \& {Gieles}}]{lamers10}
{Lamers}, H.~J.~G.~L.~M., {Baumgardt}, H., \& {Gieles}, M. 2010, \mnras, 409,
  305

\bibitem[{{Lamers} \& {Gieles}(2006)}]{lamers06a}
{Lamers}, H.~J.~G.~L.~M. \& {Gieles}, M. 2006, \aap, 455, L17

\bibitem[{{Lamers} {et~al.}(2005){Lamers}, {Gieles}, {Bastian}, {Baumgardt},
  {Kharchenko}, \& {Portegies Zwart}}]{lamers05}
{Lamers}, H.~J.~G.~L.~M., {Gieles}, M., {Bastian}, N., {et~al.} 2005, \aap,
  441, 117

\bibitem[{{Larsen}(2009)}]{larsen09}
{Larsen}, S.~S. 2009, \aap, 494, 539

\bibitem[{{Larsen} {et~al.}(2014){Larsen}, {Brodie}, {Forbes}, \&
  {Strader}}]{larsen14}
{Larsen}, S.~S., {Brodie}, J.~P., {Forbes}, D.~A., \& {Strader}, J. 2014, \aap,
  565, A98

\bibitem[{{Larsen} {et~al.}(2001){Larsen}, {Brodie}, {Huchra}, {Forbes}, \&
  {Grillmair}}]{larsen01}
{Larsen}, S.~S., {Brodie}, J.~P., {Huchra}, J.~P., {Forbes}, D.~A., \&
  {Grillmair}, C.~J. 2001, \aj, 121, 2974

\bibitem[{{Larsen} {et~al.}(2012{\natexlab{a}}){Larsen}, {Brodie}, \&
  {Strader}}]{larsen12}
{Larsen}, S.~S., {Brodie}, J.~P., \& {Strader}, J. 2012{\natexlab{a}}, \aap,
  546, A53

\bibitem[{{Larsen} {et~al.}(2012{\natexlab{b}}){Larsen}, {Strader}, \&
  {Brodie}}]{larsen12b}
{Larsen}, S.~S., {Strader}, J., \& {Brodie}, J.~P. 2012{\natexlab{b}}, \aap,
  544, L14

\bibitem[{{Leaman} {et~al.}(2013){Leaman}, {Venn}, {Brooks}, {Battaglia},
  {Cole}, {Ibata}, {Irwin}, {McConnachie}, {Mendel}, {Starkenburg}, \&
  {Tolstoy}}]{leaman13}
{Leaman}, R., {Venn}, K.~A., {Brooks}, A.~M., {et~al.} 2013, \apj, 767, 131

\bibitem[{{Lee} \& {Jang}(2016)}]{leejang16}
{Lee}, M.~G. \& {Jang}, I.~S. 2016, \apj, 822, 70

\bibitem[{{Lee} {et~al.}(2007){Lee}, {Gim}, \& {Casetti-Dinescu}}]{lee07}
{Lee}, Y.-W., {Gim}, H.~B., \& {Casetti-Dinescu}, D.~I. 2007, \apjl, 661, L49

\bibitem[{{Lemasle} {et~al.}(2008){Lemasle}, {Fran{\c c}ois}, {Piersimoni},
  {Pedicelli}, {Bono}, {Laney}, {Primas}, \& {Romaniello}}]{lemasle08}
{Lemasle}, B., {Fran{\c c}ois}, P., {Piersimoni}, A., {et~al.} 2008, \aap, 490,
  613

\bibitem[{{Lianou} {et~al.}(2010){Lianou}, {Grebel}, \& {Koch}}]{lianou10}
{Lianou}, S., {Grebel}, E.~K., \& {Koch}, A. 2010, \aap, 521, A43

\bibitem[{{Liu} {et~al.}(2011){Liu}, {Peng}, {Jord{\'a}n}, {Ferrarese},
  {Blakeslee}, {C{\^o}t{\'e}}, \& {Mei}}]{liu11}
{Liu}, C., {Peng}, E.~W., {Jord{\'a}n}, A., {et~al.} 2011, \apj, 728, 116

\bibitem[{{Livermore} {et~al.}(2015){Livermore}, {Jones}, {Richard}, {Bower},
  {Swinbank}, {Yuan}, {Edge}, {Ellis}, {Kewley}, {Smail}, {Coppin}, \&
  {Ebeling}}]{livermore15}
{Livermore}, R.~C., {Jones}, T.~A., {Richard}, J., {et~al.} 2015, \mnras, 450,
  1812

\bibitem[{{Longmore} {et~al.}(2014){Longmore}, {Kruijssen}, {Bastian}, {Bally},
  {Rathborne}, {Testi}, {Stolte}, {Dale}, {Bressert}, \& {Alves}}]{longmore14}
{Longmore}, S.~N., {Kruijssen}, J.~M.~D., {Bastian}, N., {et~al.} 2014,
  Protostars and Planets VI, 291

\bibitem[{{Lotz} {et~al.}(2001){Lotz}, {Telford}, {Ferguson}, {Miller},
  {Stiavelli}, \& {Mack}}]{lotz01}
{Lotz}, J.~M., {Telford}, R., {Ferguson}, H.~C., {et~al.} 2001, \apj, 552, 572

\bibitem[{{Mac Low} \& {Ferrara}(1999)}]{maclow99}
{Mac Low}, M.-M. \& {Ferrara}, A. 1999, \apj, 513, 142

\bibitem[{{Mackey} \& {Gilmore}(2004)}]{mackey04}
{Mackey}, A.~D. \& {Gilmore}, G.~F. 2004, \mnras, 355, 504

\bibitem[{{Mackey} {et~al.}(2010){Mackey}, {Huxor}, {Ferguson}, {Irwin},
  {Tanvir}, {McConnachie}, {Ibata}, {Chapman}, \& {Lewis}}]{mackey10}
{Mackey}, A.~D., {Huxor}, A.~P., {Ferguson}, A.~M.~N., {et~al.} 2010, \apjl,
  717, L11

\bibitem[{{Mackey} {et~al.}(2013){Mackey}, {Huxor}, {Ferguson}, {Irwin},
  {Veljanoski}, {McConnachie}, {Ibata}, {Lewis}, \& {Tanvir}}]{mackey13}
{Mackey}, A.~D., {Huxor}, A.~P., {Ferguson}, A.~M.~N., {et~al.} 2013, \mnras,
  429, 281

\bibitem[{{Madau} \& {Dickinson}(2014)}]{madau14}
{Madau}, P. \& {Dickinson}, M. 2014, \araa, 52, 415

\bibitem[{{Mannucci} {et~al.}(2009){Mannucci}, {Cresci}, {Maiolino}, {Marconi},
  {Pastorini}, {Pozzetti}, {Gnerucci}, {Risaliti}, {Schneider}, {Lehnert}, \&
  {Salvati}}]{mannucci09}
{Mannucci}, F., {Cresci}, G., {Maiolino}, R., {et~al.} 2009, \mnras, 398, 1915

\bibitem[{{Mar{\'{\i}}n-Franch} {et~al.}(2009){Mar{\'{\i}}n-Franch},
  {Aparicio}, {Piotto}, {Rosenberg}, {Chaboyer}, {Sarajedini}, {Siegel},
  {Anderson}, {Bedin}, {Dotter}, {Hempel}, {King}, {Majewski}, {Milone},
  {Paust}, \& {Reid}}]{marin-franch09}
{Mar{\'{\i}}n-Franch}, A., {Aparicio}, A., {Piotto}, G., {et~al.} 2009, \apj,
  694, 1498

\bibitem[{{Miholics} {et~al.}(2017){Miholics}, {Kruijssen}, \&
  {Sills}}]{miholics17}
{Miholics}, M., {Kruijssen}, J.~M.~D., \& {Sills}, A. 2017, \mnras~in press,
  arXiv:1705.10803

\bibitem[{{Miller} \& {Lotz}(2007)}]{miller07}
{Miller}, B.~W. \& {Lotz}, J.~M. 2007, \apj, 670, 1074

\bibitem[{{Moore} {et~al.}(2006){Moore}, {Diemand}, {Madau}, {Zemp}, \&
  {Stadel}}]{moore06}
{Moore}, B., {Diemand}, J., {Madau}, P., {Zemp}, M., \& {Stadel}, J. 2006,
  \mnras, 368, 563

\bibitem[{{Moran} {et~al.}(2014){Moran}, {Teyssier}, \& {Lake}}]{moran14}
{Moran}, C.~C., {Teyssier}, R., \& {Lake}, G. 2014, \mnras, 442, 2826

\bibitem[{{Moster} {et~al.}(2013){Moster}, {Naab}, \& {White}}]{moster13}
{Moster}, B.~P., {Naab}, T., \& {White}, S.~D.~M. 2013, \mnras, 428, 3121

\bibitem[{{Mouhcine}(2006)}]{mouhcine06}
{Mouhcine}, M. 2006, \apj, 652, 277

\bibitem[{{Mouhcine} {et~al.}(2007){Mouhcine}, {Rejkuba}, \&
  {Ibata}}]{mouhcine07}
{Mouhcine}, M., {Rejkuba}, M., \& {Ibata}, R. 2007, \mnras, 381, 873

\bibitem[{{Mouhcine} {et~al.}(2005){Mouhcine}, {Rich}, {Ferguson}, {Brown}, \&
  {Smith}}]{mouhcine05}
{Mouhcine}, M., {Rich}, R.~M., {Ferguson}, H.~C., {Brown}, T.~M., \& {Smith},
  T.~E. 2005, \apj, 633, 828

\bibitem[{{Mould} \& {Spitler}(2010)}]{mould10}
{Mould}, J. \& {Spitler}, L. 2010, \apj, 722, 721

\bibitem[{{Muratov} \& {Gnedin}(2010)}]{muratov10}
{Muratov}, A.~L. \& {Gnedin}, O.~Y. 2010, \apj, 718, 1266

\bibitem[{{Peacock} {et~al.}(2015){Peacock}, {Strader}, {Romanowsky}, \&
  {Brodie}}]{peacock15}
{Peacock}, M.~B., {Strader}, J., {Romanowsky}, A.~J., \& {Brodie}, J.~P. 2015,
  \apj, 800, 13

\bibitem[{{Peng} {et~al.}(2006){Peng}, {Jord{\'a}n}, {C{\^o}t{\'e}},
  {Blakeslee}, {Ferrarese}, {Mei}, {West}, {Merritt}, {Milosavljevi{\'c}}, \&
  {Tonry}}]{peng06}
{Peng}, E.~W., {Jord{\'a}n}, A., {C{\^o}t{\'e}}, P., {et~al.} 2006, \apj, 639,
  95

\bibitem[{{Peng} {et~al.}(2008){Peng}, {Jord{\'a}n}, {C{\^o}t{\'e}},
  {Takamiya}, {West}, {Blakeslee}, {Chen}, {Ferrarese}, {Mei}, {Tonry}, \&
  {West}}]{peng08}
{Peng}, E.~W., {Jord{\'a}n}, A., {C{\^o}t{\'e}}, P., {et~al.} 2008, \apj, 681,
  197

\bibitem[{{Portegies Zwart} {et~al.}(2010){Portegies Zwart}, {McMillan}, \&
  {Gieles}}]{portegieszwart10}
{Portegies Zwart}, S.~F., {McMillan}, S.~L.~W., \& {Gieles}, M. 2010, \araa,
  48, 431

\bibitem[{{Pota} {et~al.}(2013){Pota}, {Forbes}, {Romanowsky}, {Brodie},
  {Spitler}, {Strader}, {Foster}, {Arnold}, {Benson}, {Blom}, {Hargis},
  {Rhode}, \& {Usher}}]{pota13}
{Pota}, V., {Forbes}, D.~A., {Romanowsky}, A.~J., {et~al.} 2013, \mnras, 428,
  389

\bibitem[{{Prieto} \& {Gnedin}(2008)}]{prieto08}
{Prieto}, J.~L. \& {Gnedin}, O.~Y. 2008, \apj, 689, 919

\bibitem[{{Reina-Campos} \& {Kruijssen}(2017)}]{reinacampos17}
{Reina-Campos}, M. \& {Kruijssen}, J.~M.~D. 2017, \mnras, 469, 1282

\bibitem[{{Rejkuba} {et~al.}(2005){Rejkuba}, {Greggio}, {Harris}, {Harris}, \&
  {Peng}}]{rejkuba05}
{Rejkuba}, M., {Greggio}, L., {Harris}, W.~E., {Harris}, G.~L.~H., \& {Peng},
  E.~W. 2005, \apj, 631, 262

\bibitem[{{Rejkuba} {et~al.}(2014){Rejkuba}, {Harris}, {Greggio}, {Harris},
  {Jerjen}, \& {Gonzalez}}]{rejkuba14}
{Rejkuba}, M., {Harris}, W.~E., {Greggio}, L., {et~al.} 2014, \apjl, 791, L2

\bibitem[{{Rejkuba} {et~al.}(2009){Rejkuba}, {Mouhcine}, \&
  {Ibata}}]{rejkuba09}
{Rejkuba}, M., {Mouhcine}, M., \& {Ibata}, R. 2009, \mnras, 396, 1231

\bibitem[{{Rieder} {et~al.}(2013){Rieder}, {Ishiyama}, {Langelaan}, {Makino},
  {McMillan}, \& {Portegies Zwart}}]{rieder13}
{Rieder}, S., {Ishiyama}, T., {Langelaan}, P., {et~al.} 2013, \mnras, 436, 3695

\bibitem[{{Ryan} \& {Norris}(1991)}]{ryan91}
{Ryan}, S.~G. \& {Norris}, J.~E. 1991, \aj, 101, 1865

\bibitem[{{Saviane} {et~al.}(2000){Saviane}, {Rosenberg}, {Piotto}, \&
  {Aparicio}}]{saviane00}
{Saviane}, I., {Rosenberg}, A., {Piotto}, G., \& {Aparicio}, A. 2000, \aap,
  355, 966

\bibitem[{{Schuberth} {et~al.}(2010){Schuberth}, {Richtler}, {Hilker},
  {Dirsch}, {Bassino}, {Romanowsky}, \& {Infante}}]{schuberth10}
{Schuberth}, Y., {Richtler}, T., {Hilker}, M., {et~al.} 2010, \aap, 513, A52

\bibitem[{{Shapiro} {et~al.}(2010){Shapiro}, {Genzel}, \& {F{\"o}rster
  Schreiber}}]{shapiro10}
{Shapiro}, K.~L., {Genzel}, R., \& {F{\"o}rster Schreiber}, N.~M. 2010, \mnras,
  403, L36

\bibitem[{{Sharina} {et~al.}(2010){Sharina}, {Chandar}, {Puzia}, {Goudfrooij},
  \& {Davoust}}]{sharina10}
{Sharina}, M.~E., {Chandar}, R., {Puzia}, T.~H., {Goudfrooij}, P., \&
  {Davoust}, E. 2010, \mnras, 405, 839

\bibitem[{{Sharina} {et~al.}(2005){Sharina}, {Puzia}, \& {Makarov}}]{sharina05}
{Sharina}, M.~E., {Puzia}, T.~H., \& {Makarov}, D.~I. 2005, \aap, 442, 85

\bibitem[{{Starkenburg} {et~al.}(2010){Starkenburg}, {Hill}, {Tolstoy},
  {Gonz{\'a}lez Hern{\'a}ndez}, {Irwin}, {Helmi}, {Battaglia}, {Jablonka},
  {Tafelmeyer}, {Shetrone}, {Venn}, \& {de Boer}}]{starkenburg10}
{Starkenburg}, E., {Hill}, V., {Tolstoy}, E., {et~al.} 2010, \aap, 513, A34

\bibitem[{{Strader} {et~al.}(2004){Strader}, {Brodie}, \& {Forbes}}]{strader04}
{Strader}, J., {Brodie}, J.~P., \& {Forbes}, D.~A. 2004, \aj, 127, 3431

\bibitem[{{Strader} {et~al.}(2003){Strader}, {Brodie}, {Forbes}, {Beasley}, \&
  {Huchra}}]{strader03}
{Strader}, J., {Brodie}, J.~P., {Forbes}, D.~A., {Beasley}, M.~A., \& {Huchra},
  J.~P. 2003, \aj, 125, 1291

\bibitem[{{Strader} {et~al.}(2011){Strader}, {Romanowsky}, {Brodie}, {Spitler},
  {Beasley}, {Arnold}, {Tamura}, {Sharples}, \& {Arimoto}}]{strader11}
{Strader}, J., {Romanowsky}, A.~J., {Brodie}, J.~P., {et~al.} 2011, \apjs, 197,
  33

\bibitem[{{Streich} {et~al.}(2014){Streich}, {de Jong}, {Bailin}, {Goudfrooij},
  {Radburn-Smith}, \& {Vlajic}}]{streich14}
{Streich}, D., {de Jong}, R.~S., {Bailin}, J., {et~al.} 2014, \aap, 563, A5

\bibitem[{{Swinbank} {et~al.}(2011){Swinbank}, {Papadopoulos}, {Cox}, {Krips},
  {Ivison}, {Smail}, {Thomson}, {Neri}, {Richard}, \& {Ebeling}}]{swinbank11}
{Swinbank}, A.~M., {Papadopoulos}, P.~P., {Cox}, P., {et~al.} 2011, \apj, 742,
  11

\bibitem[{{Swinbank} {et~al.}(2012){Swinbank}, {Smail}, {Sobral}, {Theuns},
  {Best}, \& {Geach}}]{swinbank12}
{Swinbank}, A.~M., {Smail}, I., {Sobral}, D., {et~al.} 2012, \apj, 760, 130

\bibitem[{{Tolstoy} {et~al.}(2004){Tolstoy}, {Irwin}, {Helmi}, {Battaglia},
  {Jablonka}, {Hill}, {Venn}, {Shetrone}, {Letarte}, {Cole}, {Primas},
  {Francois}, {Arimoto}, {Sadakane}, {Kaufer}, {Szeifert}, \&
  {Abel}}]{tolstoy04}
{Tolstoy}, E., {Irwin}, M.~J., {Helmi}, A., {et~al.} 2004, \apjl, 617, L119

\bibitem[{{Tonini}(2013)}]{tonini13}
{Tonini}, C. 2013, \apj, 762, 39

\bibitem[{{Toomre}(1964)}]{toomre64}
{Toomre}, A. 1964, \apj, 139, 1217

\bibitem[{{Trancho} {et~al.}(2007){Trancho}, {Bastian}, {Miller}, \&
  {Schweizer}}]{trancho07}
{Trancho}, G., {Bastian}, N., {Miller}, B.~W., \& {Schweizer}, F. 2007, \apj,
  664, 284

\bibitem[{{Tremaine} {et~al.}(1975){Tremaine}, {Ostriker}, \&
  {Spitzer}}]{tremaine75}
{Tremaine}, S.~D., {Ostriker}, J.~P., \& {Spitzer}, Jr., L. 1975, \apj, 196,
  407

\bibitem[{{Tremonti} {et~al.}(2004){Tremonti}, {Heckman}, {Kauffmann},
  {Brinchmann}, {Charlot}, {White}, {Seibert}, {Peng}, {Schlegel}, {Uomoto},
  {Fukugita}, \& {Brinkmann}}]{tremonti04}
{Tremonti}, C.~A., {Heckman}, T.~M., {Kauffmann}, G., {et~al.} 2004, \apj, 613,
  898

\bibitem[{{Woodley} {et~al.}(2007){Woodley}, {Harris}, {Beasley}, {Peng},
  {Bridges}, {Forbes}, \& {Harris}}]{woodley07}
{Woodley}, K.~A., {Harris}, W.~E., {Beasley}, M.~A., {et~al.} 2007, \aj, 134,
  494

\bibitem[{{Woodley} {et~al.}(2010){Woodley}, {Harris}, {Puzia}, {G{\'o}mez},
  {Harris}, \& {Geisler}}]{woodley10}
{Woodley}, K.~A., {Harris}, W.~E., {Puzia}, T.~H., {et~al.} 2010, \apj, 708,
  1335

\bibitem[{{Zepf} \& {Ashman}(1993)}]{zepf93}
{Zepf}, S.~E. \& {Ashman}, K.~M. 1993, \mnras, 264, 611

\end{thebibliography}

\end{document}